\documentclass[]{emulateapj}
\slugcomment{{\sc Accepted to ApJ:} June 1, 2014}
\newcommand{\ms}{m\ s\ensuremath{^{-1}}}
\newcommand{\mass}{\mathcal{M}}
\newcommand{\radius}{\mathcal{R}}
\newcommand{\nsd}{\mathcal{A}}

\newcommand{\SPHIGA}{{\sc{sphiga}}}

\newcommand{\ve}[1]{\mathbf{#1}}

\newcommand{\Kepler}{\textit{Kepler}}
\newcommand{\s}[1]{\mathrm{#1}}
\newcommand{\icarus}{Icarus}

\newcommand{\figscale}{1.2}
\newcommand{\figscaletwo}{1}
\newcommand{\figscalethree}{1}

\newcommand{\figp}[1]{(\textit{#1})}
\newcommand{\pomega}{\varpi}
\newcommand{\prob}{\mathcal{P}}

\begin{document}
\shorttitle{In-situ Planet Formation}

\shortauthors{Meschiari}

\title{Circumbinary Planet Formation in the Kepler-16 System. II. A Toy Model for In-situ Planet Formation within a Debris Belt}

\author{Stefano Meschiari\altaffilmark{1}}
\altaffiltext{1}{
McDonald Observatory, University of Texas at Austin, Austin, TX 78712
}
\email{stefano@astro.as.utexas.edu}

\begin{abstract}
Recent simulations have shown that the formation of planets in circumbinary configurations (such as those recently discovered by \Kepler{}) is dramatically hindered at the planetesimal accretion stage. The combined action of the binary and the protoplanetary disk acts to raise impact velocities between km-sized planetesimals beyond their destruction threshold, halting planet formation within at least 10 AU from the binary. It has been proposed that a primordial population of ``large'' planetesimals (100 km or more in size), as produced by  turbulent concentration mechanisms, would be able to bypass this bottleneck; however, it is not clear whether these processes are viable in the highly perturbed circumbinary environments. We perform two-dimensional hydrodynamical and $N$-body simulations to show that km-sized planetesimals and collisional debris can drift and be trapped in a belt close to the central binary. Within this belt, planetesimals could initially grow by accreting debris, ultimately becoming ``indestructible'' seeds that can accrete other planetesimals in-situ despite the large impact speeds. We find that large, indestructible planetesimals can be formed close to the central binary within $10^5$ years, therefore showing that even a primordial population of ``small'' planetesimals can feasibly form a planet.
\end{abstract}

\section{Introduction}
The current paradigm of planet formation includes a stage where solid bodies (``planetesimals'') that have become large enough start decoupling from the aerodynamic drag provided by the gas disk and interacting with each other via gravity \citep[see e.g.][for reviews]{Lissauer93, Armitage13, Youdin12}. This regime traditionally comprised objects that have a radius of order of a few kilometers, as predicted by the classical instability scenarios \citep[e.g.][]{Goldreich73}. 

The role of km-sized planetesimals as the fundamental building blocks of planetary cores is, at best, precarious for the newly discovered circumbinary planets \citep{Doyle11, Welsh12, Orosz12, Orosz12a, Schwamb12, Kostov13, Kostov14}.  Indeed, simplified simulations of planetesimal accretion have shown that the interplay between the gravitational perturbations of the stellar binary and the aerodynamic drag of the gaseous disk can inhibit planet formation in the inner few AUs of the protoplanetary disk \citep[e.g.][]{Moriwaki04, Scholl07, Meschiari12}. \citet[][hereafter M12]{Meschiari12} investigated planetesimal dynamics in the Kepler-16 system and found that planetesimal collisions inside at least 4 AU ($\approx 20 a_\s{B}$, where $a_\s{B}$ is the binary semi-major axis) were largely destructive, due to the high collisional velocities. We proposed that Kepler-16 b formed in the outer regions of the disk, and subsequently migrated to its currently observed location ($\approx 3 a_\s{B}$), likely via tidal interaction with the gas disk \citep{Pierens07, Pierens08}. Subsequently,  \citet{Meschiari12c} found that turbulent density fluctuations can potentially further increase impact speeds, pushing the accretion-friendly region out to 10 AU \citep[although a more sophisticated treatment of the stochastic torques might be warranted;][]{Okuzumi13}.

The analysis of M12, however, assumed a static, axisymmetric gas disk as the source of aerodynamic drag for the sake of computational expediency. Simulations that included the hydrodynamical evolution of the gas disk suggested an even more disturbed environment, due to the development of bulk eccentricity and spiral perturbations in the disk \citep[e.g.][]{Paardekooper08, Marzari12, Muller12, Pelupessy12}. Recently, \citet{Marzari13} self-consistently followed the hydrodynamical evolution of the disk together with a swarm of planetesimals (with radii of 5 and 25 km). This setup allowed them to record a large number of  planetesimal impact events throughout the disk. The inferred impact velocities (on the order of 100-1000 \ms) were beyond the critical value for planetesimal destruction. Therefore, they concluded that embryo formation is inhibited everywhere inside about 10 AU for km-sized planetesimals, again suggesting the planetary core is formed far from the binary. The migration scenario is not without its downsides, however.  Chiefly, it requires a substantial amount of migration of the planet from the outer disk to its current location \citep[either through gas torques, or, less likely, driven by a fossil planetesimal disk, e.g.][]{Gong13}. Additionally, although impact speeds are reduced below the critical value for destruction outside 10 AU, they will still be high enough to preclude runaway growth; therefore, even if planetesimals are able to accrete, they will do so at a slower rate than that traditionally assumed in single-star environments. 

Alternatively, in-situ formation at the inferred collision speeds could be possible if we allowed for rapid formation of planetesimals at least 100-300 km (depending on the material strength of the planetesimals) in size, according to the destruction criterion of \citet{Stewart09}. Such large planetesimals would be essentially indestructible and accrete in spite of the large impact velocities. Recent planetesimal formation theories actually seem to favor the direct formation of massive bound clumps (between a fraction to several times the mass of Ceres) through the streaming instability \citep{Johansen07, Johansen11}, lending credence to the latter scenario.  However, it is not clear whether the results of the numerical simulations of the streaming instability are applicable to the circumbinary context. Firstly, the instability is modeled numerically within a corotating shearing box, which is assumed to be representative of the disk. This approximation is manifestly inaccurate in the binary environment, which is subject to time-dependent perturbations from the central binary (which in turn will excite non-axisymmetric, time-dependent perturbations in the gas disk). Therefore, it might be premature to expect that this mechanism can operate in the same manner within the circumbinary environment. Secondly, meter-sized boulders (the building blocks that concentrate and gravitationally collapse in overdense regions) are expected to collide at very high speeds and fragment due to the binary perturbations (regardless of the amplitude of stochastic turbulent torques, which is the main limiting factor in the aforementioned simulations). Finally, although direct formation of large planetesimals appears to be bolstered by the properties of the asteroids census in the Solar System \citep[e.g.][]{Morbidelli09}, other authors disagree \citep[e.g.][]{Weidenschilling11, Fortier13}. It is fair to state that at present, there is no consensus on how planetesimals form is in the circumbinary environment, including an appropriate initial size spectrum.

Although the results of \citet{Marzari13} appear to preclude in-situ planet formation from km-sized planetesimal, their simulations have a number of limitations (chiefly due to the high computational requirements of their approach). The simulations only spanned a limited time interval ($2\times 10^4$ years), and did not allow for planetesimal accretion and fragmentation. Therefore, they were restricted to measuring impact speeds and deeming the disk inside 10 AU accretion-unfriendly. The evolution of the planetesimal size spectrum, in particular, can change the boundary of the accretion-friendly zone, as shown by \citet{Paardekooper12} (hereafter P12). In particular, if a fraction of planetesimals could survive collisional grinding and acquire an alternative pathway to growing large enough to become indestructible ($\radius_\s{pl} > 100$ km), then planet formation could proceed even at the significant impact speeds found close to the central binary.

Radial drift could be potentially helpful to this goal. Solid objects in Keplerian orbits will lose angular momentum due to the aerodynamic drag acceleration \citep{Adachi76, Weidenschilling77}:
\begin{equation}
\ve{F}_\s{drag} = - f(v_\s{rel}, c_\s{s})\ \frac{\rho_\s{g}}{\rho_\s{pl} \radius_\s{pl}}\ \ve{v}_\s{rel} \ ,
\label{eqn:drag}
\end{equation}
where $\rho_\s{g}$ and $\rho_\s{pl}$ are the gas and plantesimal density, respectively, $c_\s{s}$ is the local sound speed and $\ve{v}_\s{rel} = \ve{v}_\s{gas} - \ve{v}_\s{pl}$ is the relative velocity between the gas streamline and the planetesimal. The specific form of the function $f$ (i.e. whether the Epstein or quadratic regime is appropriate) will depend on the radius of the body $\radius_\s{pl}$.  The gas speed is given by
\begin{equation}
v_\s{gas}^2 = \frac{\mass_\s{*}}{R} + \frac{R}{\Sigma} \frac{dP}{dR} = \Omega^2 R^2 \left[ 1 - h^2 + \frac{h^2 R}{\Sigma}\ \frac{d\Sigma}{dR} \right] \ ,\label{eqn:gasspeed}
\end{equation}
where $\mass_* = \mass_1+\mass_2$ is the total binary mass, $P$ and $\Sigma$ are the local pressure and surface density, respectively, and $\Omega = R v_K$ is the circular angular speed (in units where $G=1$). The second equality holds for a disk with constant aspect ratio $h$ ($h = H/R$, where $H$ is the local scale height) and an isothermal equation of state $P = \Sigma h^2 R^2 \Omega^2$. For typical surface density profiles (with $d\Sigma/dR < 0$), the gas speed is lower than the local Keplerian speed, resulting in the solid body experiencing a headwind and spiraling inwards. While the surface density gradient can be assumed to be negative in single stars environments (until very close to the central star), the presence of a central binary will truncate the disk starting at a specific radius \citep[$\approx 3 a_\s{b}$;][]{Artymowicz94}, resulting in a density gradient inversion and a local pressure maximum. At this locus, solid bodies could be ``trapped'', since the drag force will change sign across the pressure maximum (vanishing at the pressure maximum). Therefore, solids will tend to accumulate at the boundary between super-Keplerian/sub-Keplerian gas speeds, as seen in numerical simulations that include a non-uniform pressure profile \citep[e.g.][]{Haghighipour03, Fouchet07, Kretke09, Kato10}. This trapping was also observed by \citet{Marzari08}, who remarked that, in their simulations, small bodies (100 m in size) concentrated in the inner regions of the disk surrounding a binary star.

Planetesimals are also liable to accumulating in the same region, although on a much longer timescale than meter-sized debris. The drifting speeds for a swarm of km-sized planetesimals in low-eccentricity orbits embedded in an axisymmetric disk (the typical situation in single-star environments) are typically very low (of order $10^{-6}$ -- $10^{-7}$ AU/yr, depending on the planetesimal location).  However, in the setup considered in this paper the central binary will force a large eccentricity on the planetesimals, such that the velocity differential $v_\s{rel}$ is increased, enhancing drift speeds by an order of magnitude or more (as we will show in Section \ref{sec:drift}).  This implies that planetesimal drift could be significant on the timescales relevant to planet formation. In particular, any planetesimals that survive grinding will migrate and stop at the trapping locus.

The radial drift of debris and planetesimals suggests an alternative to the migration of the planetary core scenario. Assuming that planetesimals are formed rapidly throughout the disk, then the following sequence creates an accretion-friendly planetesimal belt in the inner disk where lucky ``survivors'' might thrive and grow:
\begin{enumerate}
\item A majority of planetesimals undergo rapid grinding throughout the disk, creating a large amount of small debris (``dust'').
\item The dust created by planetesimal grinding quickly drifts into the pressure maximum, and persists there (it cannot drift further in due to the combined action of the aerodynamic drag and the angular momentum barrier).
\item Planetesimal grinding continues until collision timescales become longer than drift timescales. Planetesimals that survived fragmentation can therefore migrate and stop into the debris belt.
\item Within the belt, planetesimals can grow by sweeping up dust \citep[a localized version of the ``snowball model'' of ][]{Xie10}. Due to the ongoing accumulation of debris within the belt, then planetesimals might quickly become large enough to be indestructible. These bodies may become ``seeds'' for runaway growth and start accreting other planetesimals, ultimately accumulating into a single planetary core. Further planetesimal drift extend the feeding zone of the core.
\end{enumerate}
\begin{figure}
\epsscale{\figscale}
\plotone{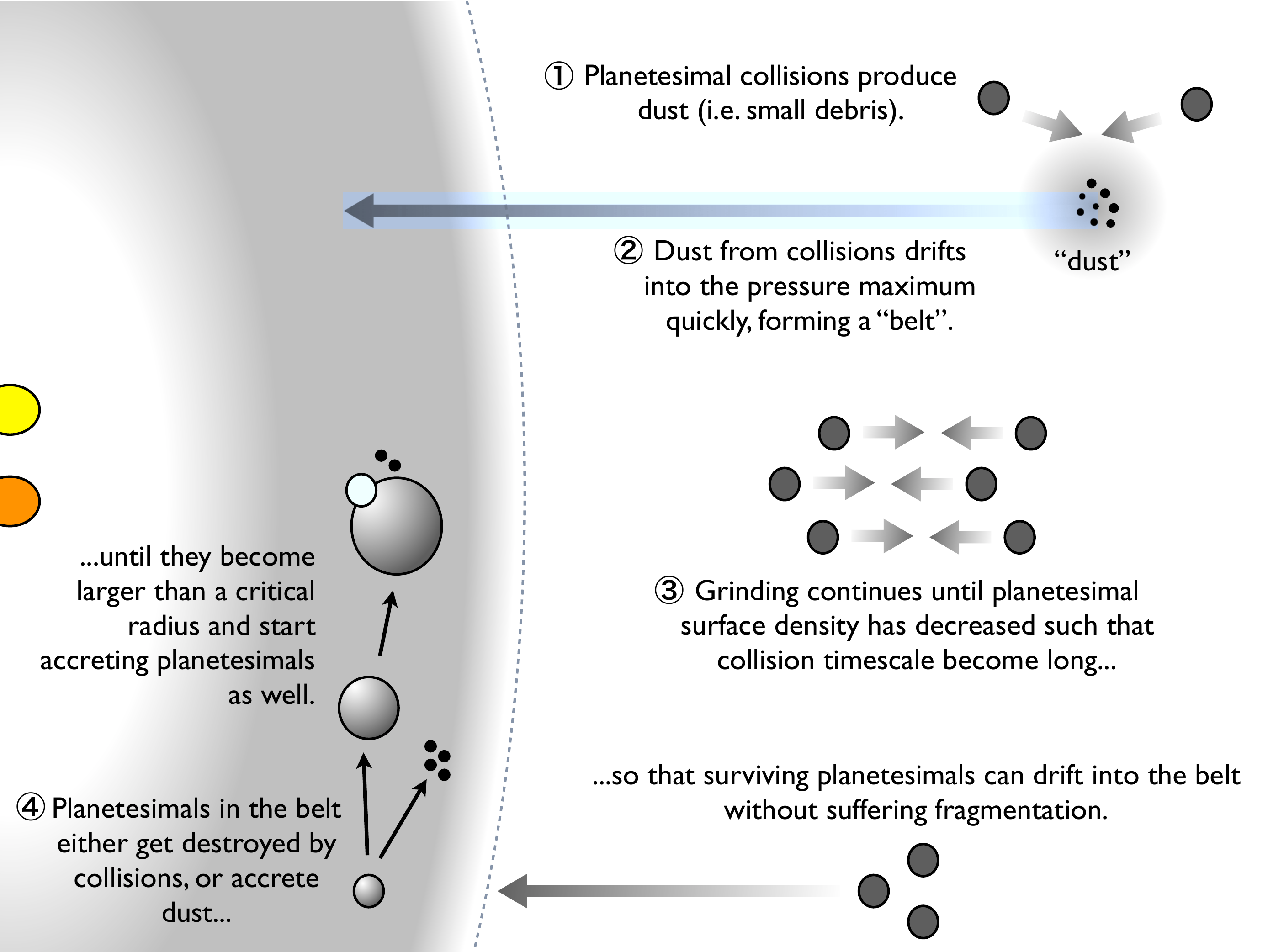}
\caption{A sketch of the destruction-drift-reaccumulation model. Each step is further described in the text, as numbered.}\label{fig:sketch}
\end{figure}

We sketch this process in Figure \ref{fig:sketch}. Although this scenario might seem plausible at face value, the arguments presented so far still represent a substantial simplification of the complex physical setup. Indeed, both the gas disk and the swarm of planetesimals will develop significant eccentricity and precession due to the gravitational perturbations of the central binary. Therefore, it is not immediately clear that planetesimals and debris could reside in a well-confined belt, as hypothesized above. 

This paper attempts to ascertain whether the ``destruction-drift-reaccumulation'' process sketched above can be physically feasible in a realistic circumbinary environment. We developed a numerical approach that, despite numerous simplifications, strives to model the various physical processes at work for the long timescales involved (up to about $10^5$ years)  as faithfully as possible. We discuss our code and planetesimal dynamics in Section \ref{sec:drift}. In Section \ref{sec:results}, we build from the results of Section \ref{sec:drift} and create a toy model that follows planetesimals growth in the belt. Finally, we conclude in Section \ref{sec:conc}.

\section{Planetesimal drift}\label{sec:drift}
To validate the scenario proposed in the Introduction, we first need to verify that planetesimals will halt their migration close to the pressure maximum. This will require a more sophisticated approach than that employed in M12, which assumed an axisymmetric and static gas background. On the other hand, given that we aim to model planetesimal dynamics for long timescales, the self-consistent simulations of \citet{Marzari13} would be exceedingly expensive. Therefore, we decided to take a hybrid approach, whereby we first run high-resolution, two-dimensional hydrodynamical simulations for $10^4$ years. We then evaluate the output of the simulations in order to derive a highly simplified model of the gas dynamics. This model allows us to provide a more realistic approximation for the aerodynamic drag and a gravitational term arising from the disk, coupled with our $N$-body code. As shown below, an eccentric, rigidly precessing disk is a reasonable (if crude) approximation to the full hydrodynamical output. Finally, we run our $N$-body code fitted with the gas model to verify whether planetesimals are trapped close to the truncation radius.

\subsection{Hydrodynamical simulations}\label{sec:fargo}

\begin{figure}
\epsscale{\figscalethree}
\plotone{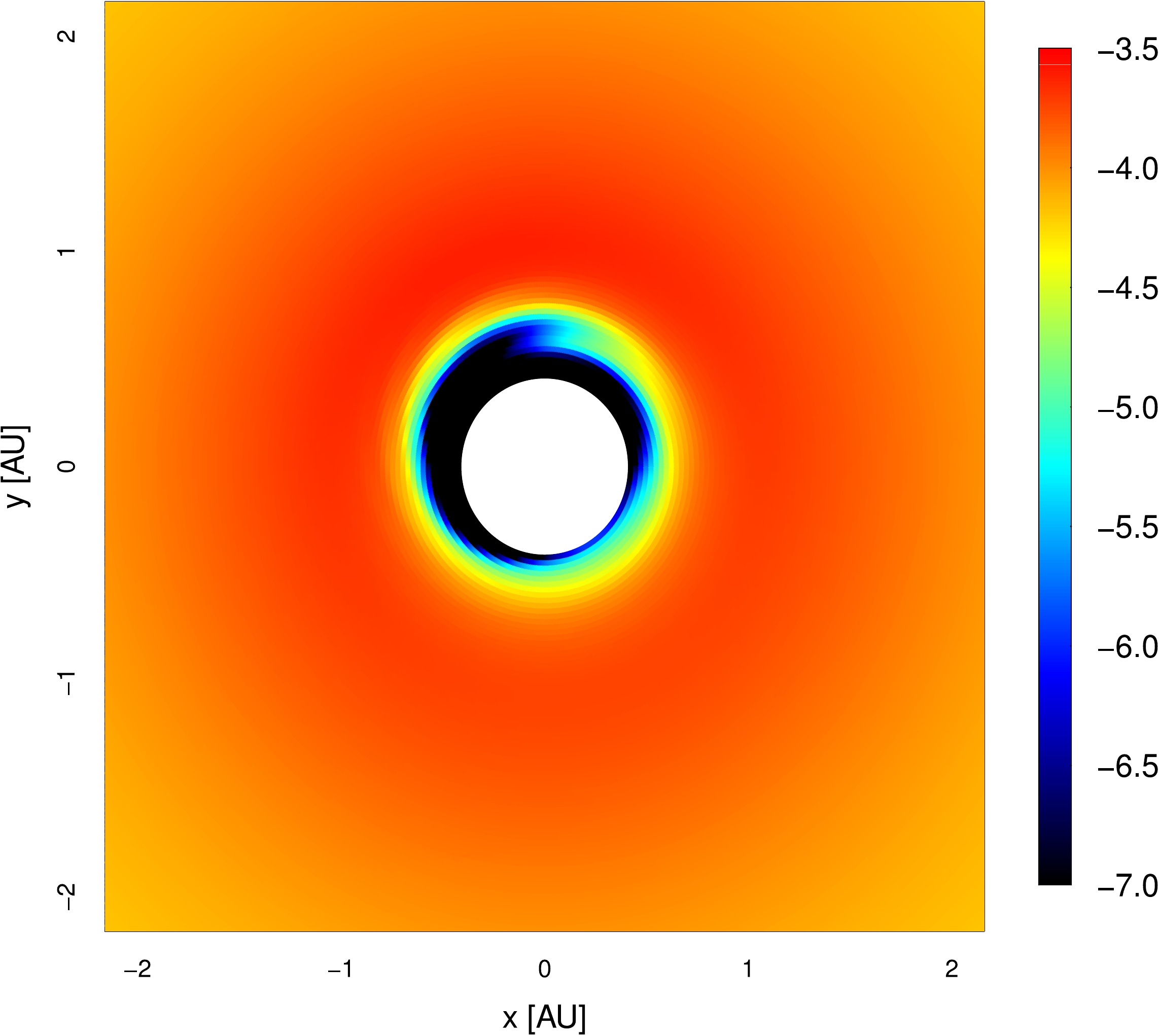}\\
\plotone{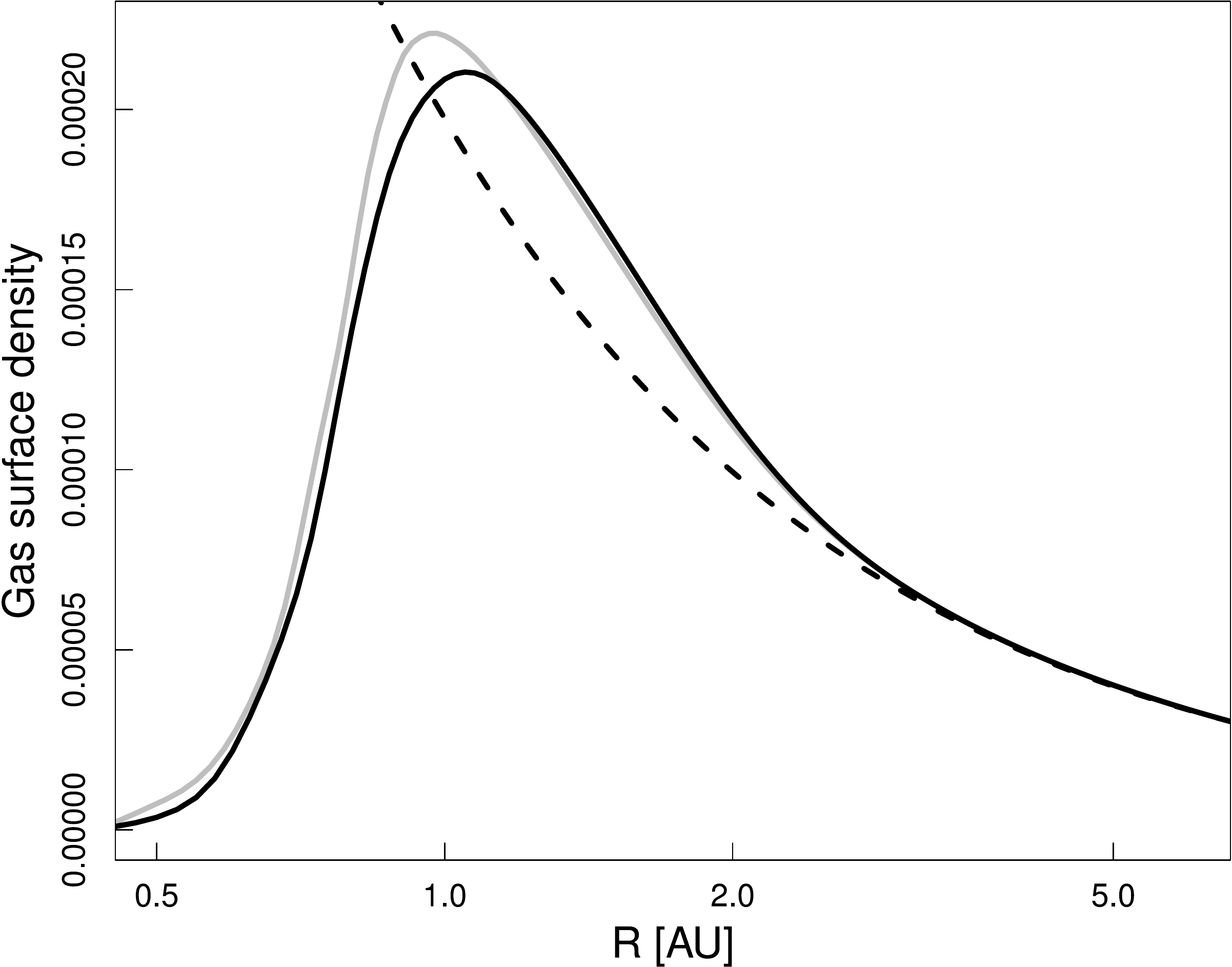}\\
\plotone{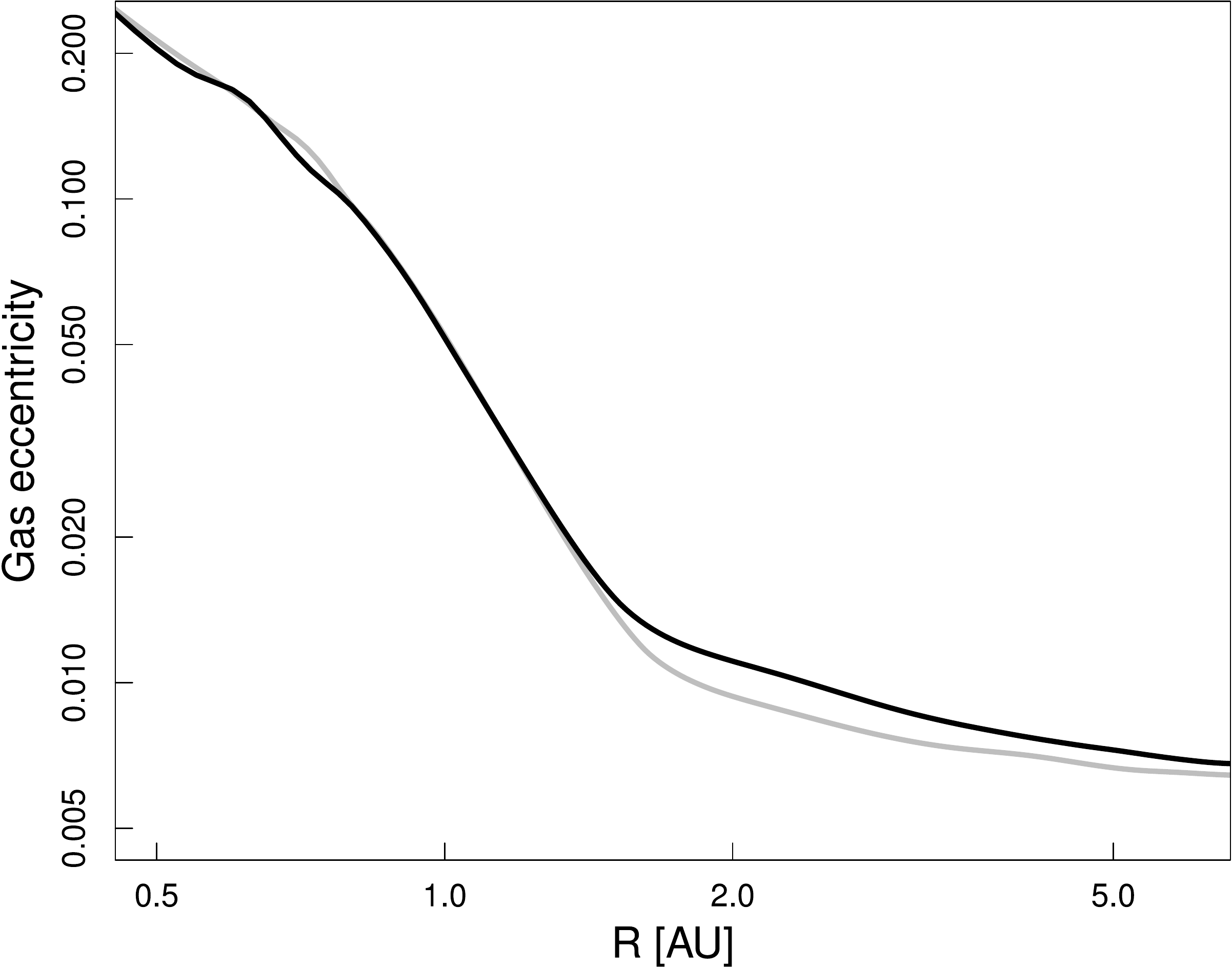}
\caption{Snapshot of the gas disk after $10^4$ years. Dimensionful quantities are expressed in terms of code units. \figp{Top} Logarithmic surface density. \figp{Middle} Azimuthally averaged surface density as a function of the distance from the center of mass (the dashed line represents the initial surface density profile). \figp{Bottom} Azimuthally averaged eccentricity. The grey line represents a higher-resolution run (with $N_\s{\theta} = 768$ and $N_\s{r} = 512$).}\label{fig:hydro}
\end{figure}

\begin{figure}
\epsscale{\figscaletwo}
\plotone{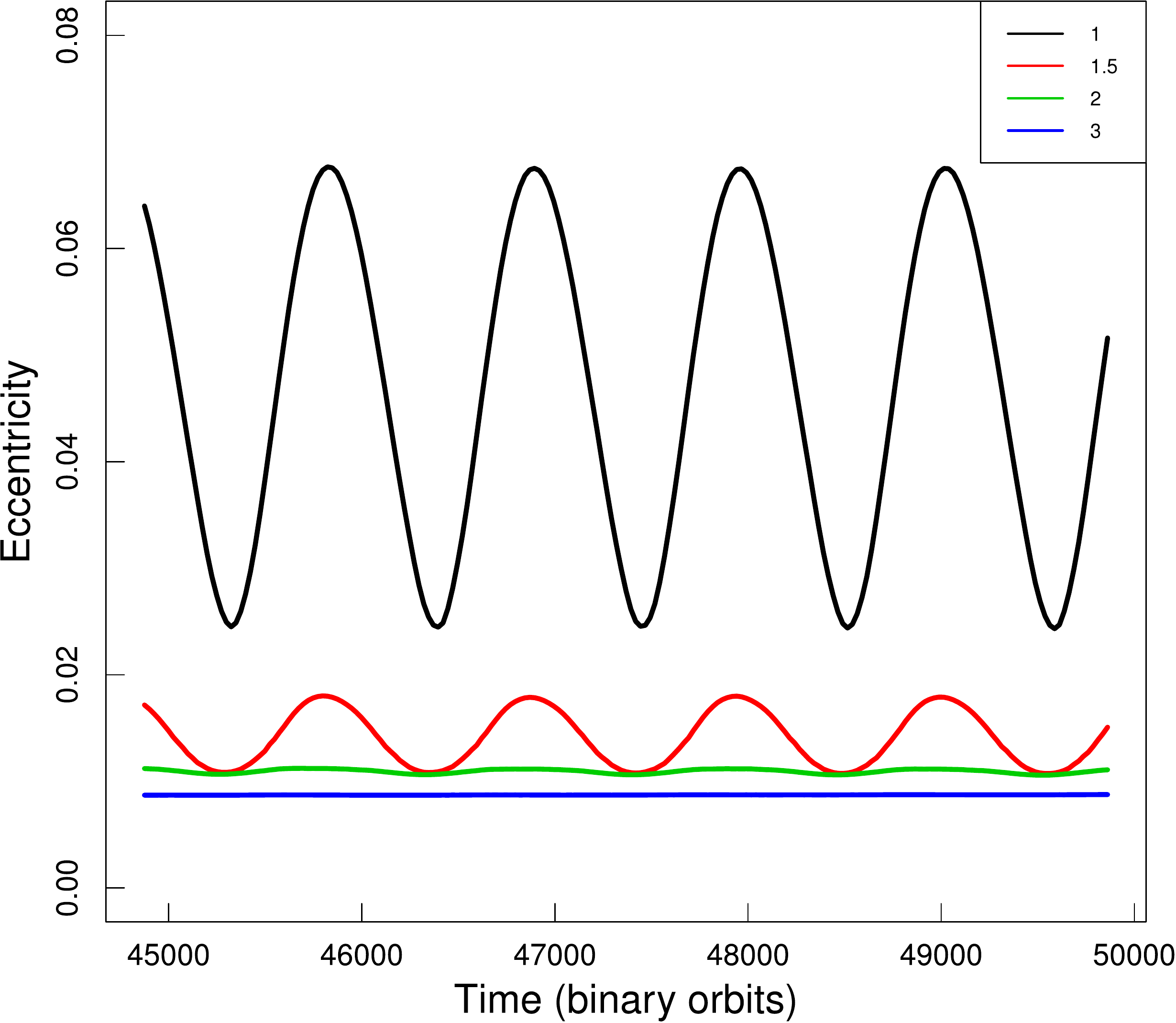}
\plotone{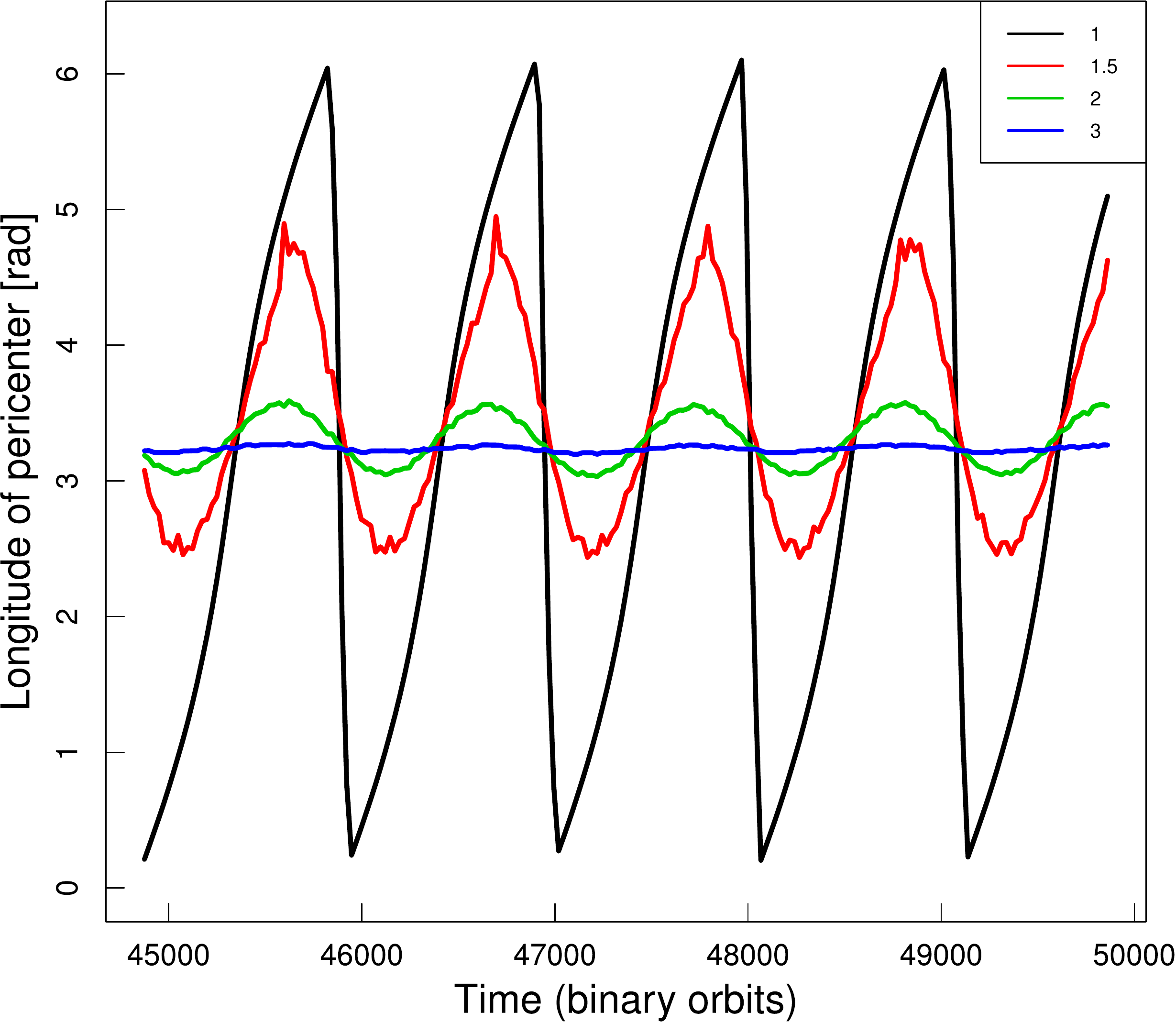}
\caption{Sample evolution of the gas eccentricity and longitude of pericenter at a few representative radii (expressed in AU).  The gas eccentricity oscillates as a function of time, although the average eccentricity is approximately constant at each radii. The inner disk (inside $\approx$ 1.5 AU) librates with a large amplitude through almost the full 2$\pi$, while the outer disk librates with a smaller amplitude. }\label{fig:hydro_byr}
\end{figure}

We ran a host of two-dimensional hydrodynamical simulations using the isothermal FARGO code \citep{Masset00}. We modified the code to work in barycentric coordinates. For the rest of this paper, we consider the binary parameters of the Kepler-16 system ($\mass_1 = 0.69 \mass_\sun$,  $\mass_2 = 0.2 \mass_\sun$, $P_\s{B} = 41$ days, $e_\s{B} = 0.15$) for easy comparison with previous works. The Keplerian orbital elements of the binary are fixed.

We employ a simple thin disk model as our initial configuration. We assume a disk with surface density $\Sigma \propto R^{-1}$ \citep[with the same normalization as the minimum-mass solar nebula;][]{Hayashi81} and constant aspect ratio $h = 0.05$; the disk is initially circular. The disk is modeled by a grid with $N_\theta =$ 512 azimuthal divisions and $N_\s{r} = $ 384 radial divisions (with an arithmetic spacing). The radial direction covers between 0.4 AU and 10 AU and ends with an open boundary on either side of the grid. We added an exponential taper to the surface density close to the inner and outer boundary. This setup is similar to that of \citet{Marzari08}, although we take a different slope for the surface density and we neglect the self-gravity of the disk. Finally, we assume a constant kinematic viscosity for the disk ($\nu = 3\times 10^{-6}$, corresponding to $\alpha \approx 10^{-3}$ at 5 AU).

We evolve our disk model for 10,000 years (more than $10^5$ binary orbits). The disk very quickly opens an inner cavity (within a few tens of binary orbits). After an initial period of activity near the cavity, the surface density evolves very little throughout the simulation, except for a slow expansion of the inner cavity \citep[with the surface density maximum moving outwards, as expected for a viscous external disk;][]{Pringle91}. The evolution surface density is shown in the top panels of Figure \ref{fig:hydro}. 

Subsequently, we calculated the semi-major axis of each cell and averaged the eccentricity in each semi-major axis bin (Figure \ref{fig:hydro}, bottom panel). The eccentricity is calculated from the velocity components of each grid cell. The eccentricity peaks close to the inner cavity and rapidly declines, such that the outer disk (outside $\approx 1.5$ AU) has negligible eccentricity. We find this situation to be qualitatively similar to the results of \citet{Pelupessy12}, but quite different from the results of \citet{Marzari13}. The latter simulations included more realistic disk thermodynamics than our simple isothermal assumption. The simulation was run at a higher resolution as well (with $N_\theta = 768$ and $N_\s{r} = 512$ grid points) in order to investigate the sensitivity of the numerical results to the grid resolution. We found that the output of our higher-resolution runs are consistent with our standard-resolution runs, aside from small differences in the disk eccentricity and surface density close to the inner edge (Figure \ref{fig:hydro}).

Figure \ref{fig:hydro_byr} shows the eccentricity and longitude of periastron evaluated at a few select radii. Although the eccentricity has an average value that is approximately constant throughout the simulation, there are substantial periodic oscillations. The inner disk (inside $\approx 1.5$ AU) appears to be librating with a very large amplitude and a period of about 1,000 binary orbits (almost to the point of rigid precession), whereas the outer disk appears to be librating around $\varpi \approx \varpi_\s{b} = \pi$.

We note that the eccentricity thus found is derived assuming a Keplerian potential around a body placed in the center of mass of the binary with mass $\mass_*$ (i.e., such that the potential is $U = G\mass_*/r$). This might add a substantial eccentricity in the initial conditions of the disk, due to the fact that the azimuthal velocity of the fluid cells are started at a lower speed than a circular orbit in the true axisymmetric potential \citep{Rafikov13}. The fictitious eccentricity does not modify the overall dynamics, as the orbital elements of the disk influence the planetesimal dynamics through Equation \ref{eqn:drag} only; this means that the orbital elements are converted back into a physical velocity vector at each location within the disk (i.e., the orbital elements are just a convenient intermediate representation). However, it does imply that the disk is not initialized in a ``true'' circular state. For the specific case of the Kepler-16 system, the fictitious eccentricity should be of the order of $e(r) \approx 7\times 10^{-3}\times r^{-2}$ (with $r$ expressed in AU, using Formula 26 of \citet{Rafikov13}). 

\subsection{Model setup}\label{sec:model_setup}
Ideally, we would like to follow planetesimal drift self-consistently by coupling test particles (planetesimals) to the hydrodynamical evolution of the disk.  However, as we will see in the next section, this approach is prohibitively expensive since, despite the comparatively rapid drift experienced in the inner 2 AU of the disk, the relevant timescales are much longer than those modeled above (up to a factor of 100).  Therefore, it is desirable, if not entirely self-consistent, to model the gas evolution in an approximate fashion, such that the evolution of a large population of planetesimals can be followed for at least $10^6$ years.

We follow the spirit of the formulation of \citet{Beauge10} by assuming that the gas disk can be represented as an eccentric, rigidly precessing disk. We assume that each hydrodynamical quantity $q$ (i.e. surface density, sound speed and scale height) is constant on each eccentric streamline $\ve{r}$, i.e. $q = q[\ve{r}(a_\s{g}, e_\s{g}, \varpi_\s{g})]$, where $a_\s{g}$ is the gas semi-major axis, $e_\s{g}$ is the gas eccentricity and $\varpi_\s{g}$ is the longitude of periastron. The eccentricity at each semi-major axis is assumed constant for the duration of the simulation (Figure \ref{fig:hydro}), while the longitude of pericenter is assumed to be precessing with a constant period $P_\varpi$, such that $\varpi_\s{g} = 2\pi/P_\varpi\times t$ (although the outer disk is librating in our simulation, the eccentricity there is small enough to have a negligible effect on the planetesimal dynamics). Finally, we assume that the gas speed at each eccentric streamline is given by
\begin{equation}
\ve{v}_\s{g}(\ve{r}) = \ve{v}_\s{K}(\ve{r}) \times \left[1 - \xi(a_\s{g})\right] \ ,
\end{equation}
where $\xi(a_\s{g}) = h^2 - h^2 a \Sigma'/\Sigma$ (as in Equation \ref{eqn:gasspeed}) and $v_K(\ve{r})$ is the Keplerian orbital speed along the streamline. Therefore, in our model the speed differential is determined by the local surface density gradient \citep[as opposed to being constant, an assumption of the simplified model of][]{Beauge10}.  The disk model is then completed by specifying a surface density profile $\Sigma(\ve{r})$, an eccentricity profile $e_\s{g}(a_\s{g})$ and the precession period $P_\pomega$.

Once the model is specified as above, we bin the hydrodynamical quantities over a 2D grid (with 512 radial and 512 azimuthal zones) for computational convenience. This step allows us to efficiently compute hydrodynamical quantities local to the planetesimal, and optionally, the gas potential. 

Planetesimals are represented as a swarm of non-interacting test particles, subject to the sum of the gravitational force of the binary system, the aerodynamic drag and the gas potential. The aerodynamic drag is given by Equation \ref{eqn:drag}, with $f$ taking the following form in the quadratic regime:
\begin{equation}
f = \frac{3}{8} C \left|\ve{v}_\s{rel}\right| \ ,
\end{equation} 
(where the coefficient $C \approx 0.4$ for spherical bodies). The hydrodynamical quantities ($\rho_\s{g}, \ve{v}_\s{rel}, H$) are efficiently interpolated at the planetesimal location (using bilinear interpolation). The planetesimal orbits are evolved forward in time using the \SPHIGA{} code \citep{Meschiari12, Meschiari12c}, which employs an 8-th order Runge-Kutta integration scheme.

The model described above makes a number of approximations for the sake of computational expediency. The most important limitation is the choice of an isothermal equation of state for our hydrodynamical simulations, which appears to effect the gas dynamics. \citet{Marzari13} finds that adding a realistic energy equation to the hydrodynamical model results in a much more active disk, such that the eccentricity profile and potential change rapidly as a function of time. Taking these effects into account in our model may be difficult. However, it is possible that these time-dependent variations might be transient and become less important over longer timescales.

Our model assumes that the local hydrodynamical properties of the disk can be usefully represented by assuming they are constant over an eccentric streamline.  This assumption simplifies an inherently two-dimensional problem (the evolution of a hydrodynamical quantity $q[R, \phi]$) into a one-dimensional problem ($q[a_g, e_g(a_g)]$). However, \citet{Statler01} showed that this approximation is unwarranted if the disk eccentricity varies as a function of semi-major axis ($e' \ne 0$). This is exactly the case in our hydrodynamical simulation (see the bottom panel of Figure \ref{fig:hydro}). At a given semi-major axis, there is substantial scatter in $\Sigma$ and other quantities, especially close to the central binary (i.e. $\lesssim 1.5$ AU) where the perturbations are strongest. Therefore, the one-dimensional approximation presented in this Section (and further developed into a model in Section \ref{sec:evolving} and \ref{sec:results}) is not strictly correct.

We will summarize a number of caveats related to our simplified treatment of the background gas disk in Section \ref{sec:caveats}.

\subsection{Fixed axisymmetric background}
\begin{figure}
\epsscale{\figscaletwo}
\plotone{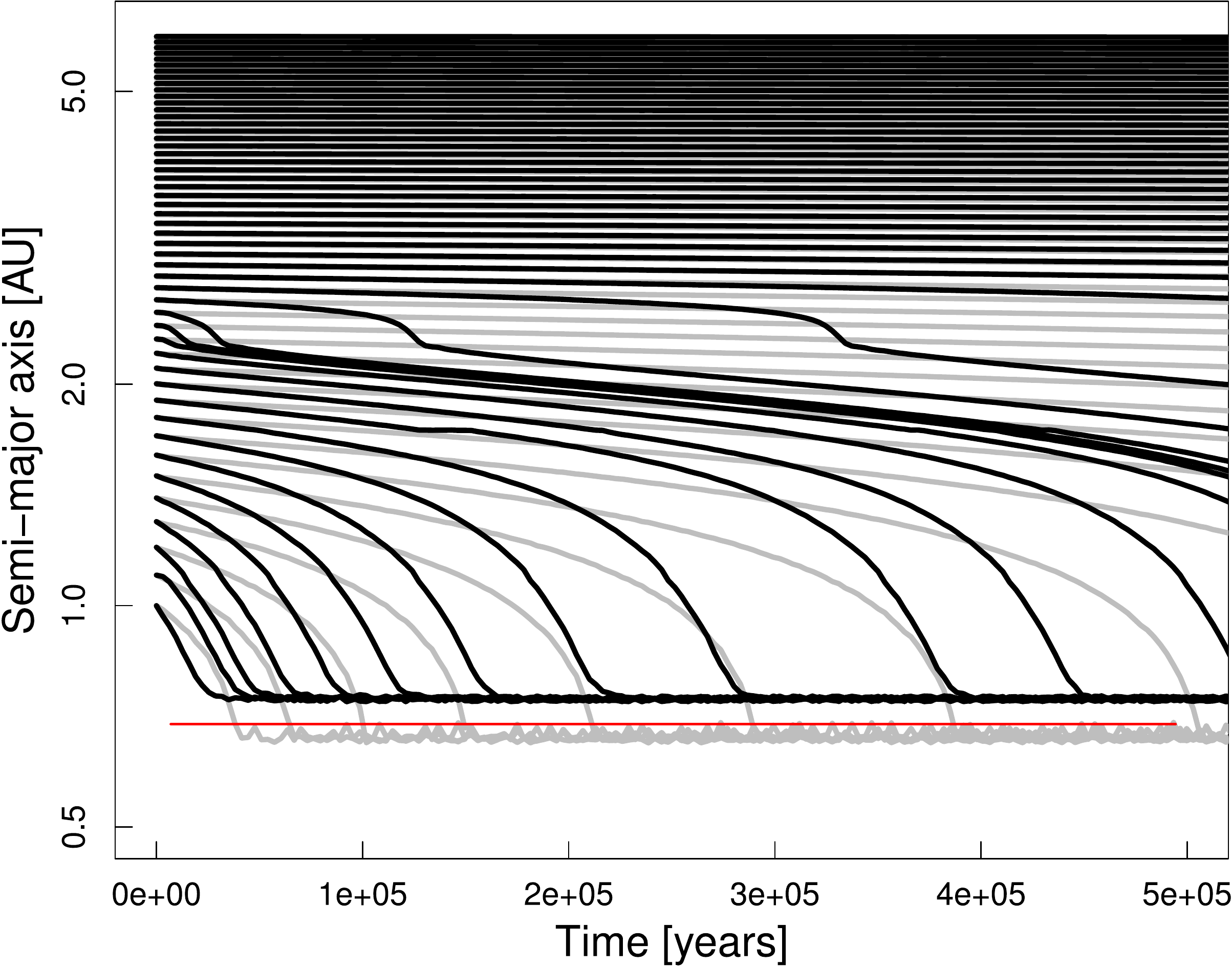}
\plotone{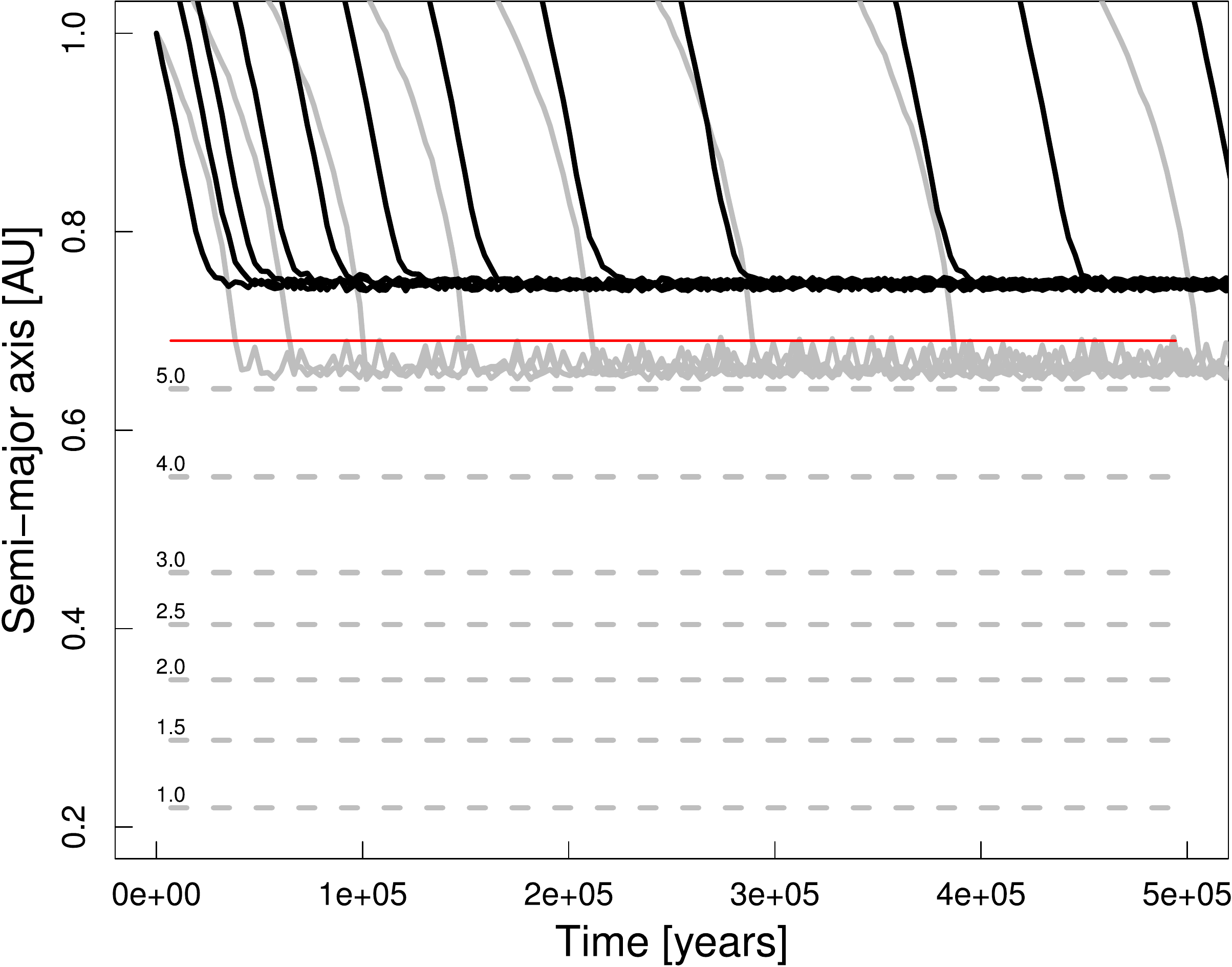}
\caption{Planetesimal drift over $5 \times 10^5$ years, in the fixed surface density gas model (with gas potential; black lines) and the axisymmetric gas model of M12 (grey lines). In the axisymmetric model, planetesimals get locked in resonance with the central binary close to the 5:1 resonance. In the fixed density model, planetesimals stop further out close to the pressure maximum. The red line represents the current location of Kepler-16 b. The bottom panel zooms in the inner AU of the disk, and sketches some representative period commensurability's with the central binary (dotted horizontal lines; each commensurability is labeled according to the ratio between the orbital period at the semi-major axis and the binary period). Notice that once planetesimals are trapped, there is very little semi-major axis variation.}\label{fig:drift}
\end{figure}

We first show planetesimal drift as computed assuming the fixed, axisymmetric gas background of M12. Figure \ref{fig:drift} shows the planetesimal drift (semi-major axis as a function of time) for planetesimals with radius $\radius_\s{pl} = 2.5$ km. The planetesimal disk inside $\approx 2$ AU is depleted within $10^6$ years. The drift speed is larger than that expected in single-star systems, due to the large eccentricity excited in the planetesimal disk \citep{Meschiari12}. 

Planetesimals do not migrate all the way to the instability limit \citep{Holman99}, however. The presence of the central binary dictates that as they migrate inwards, they are liable to being captured in a resonance with the stars \citep[analogously to the resonant capture of migrating planets; e.g.][]{Lee02}. In our case, the semi-major axis is fixed; therefore, planetesimals will  converge to a fixed location. For the set of parameters considered in this paper, planetesimals are captured just outside the 5:1 resonance \citep[where the planet also resides;][]{Popova13}.

\subsection{Fixed surface density model}

\begin{figure}
\epsscale{\figscale}
\plotone{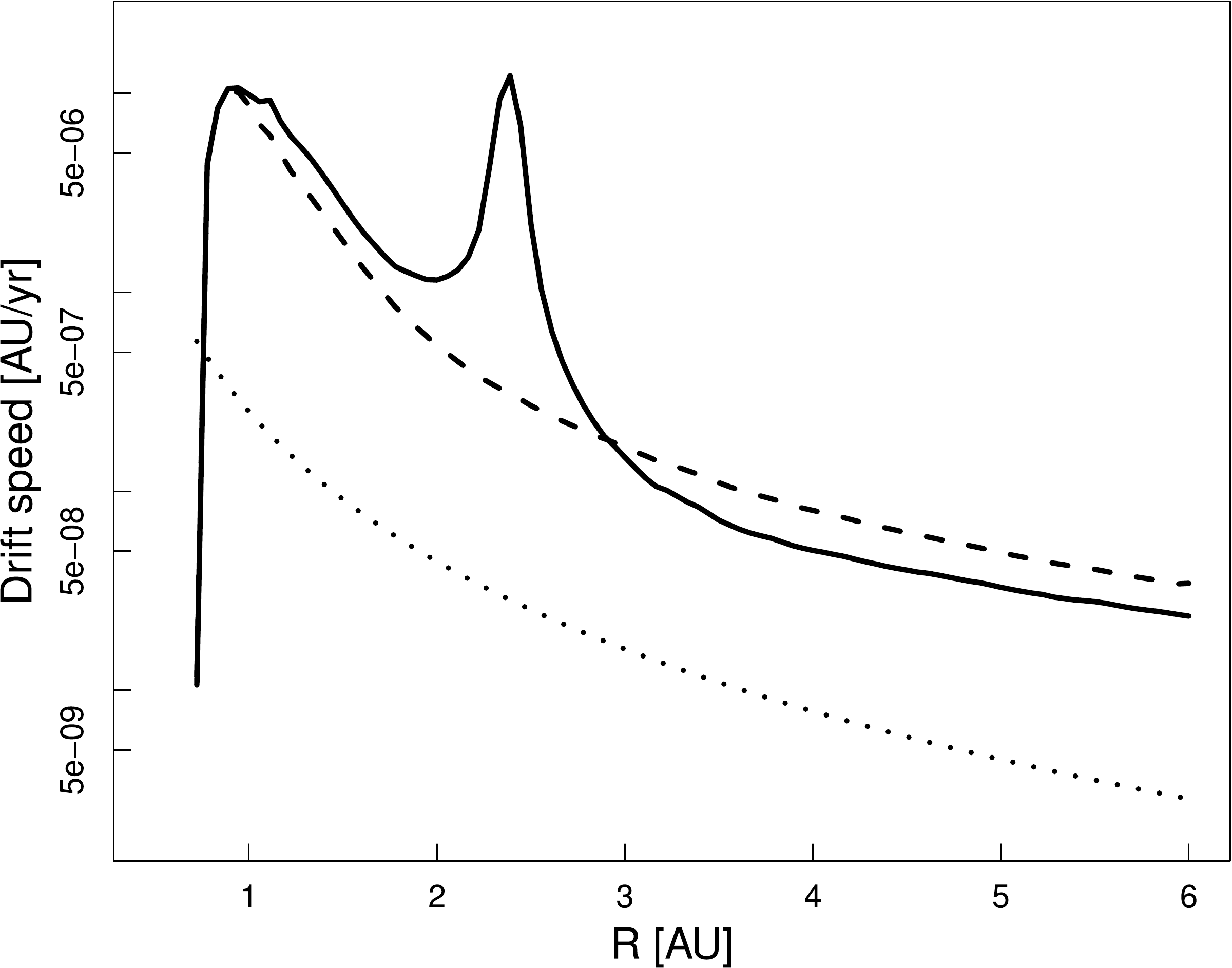}
\caption{Comparison between the drift speed for the fixed surface density model with (solid line) and without (dashed line) the gas potential, for planetesimals with $\radius_\s{pl} = 2.5$ km. Notice the large spike in drift speed at $R \approx 2$ AU. We also plot the drift speed in a single-star environment for reference (dotted line, assuming zero eccentricity).}\label{fig:vr}
\end{figure}

\begin{figure}
\epsscale{\figscaletwo}
\plotone{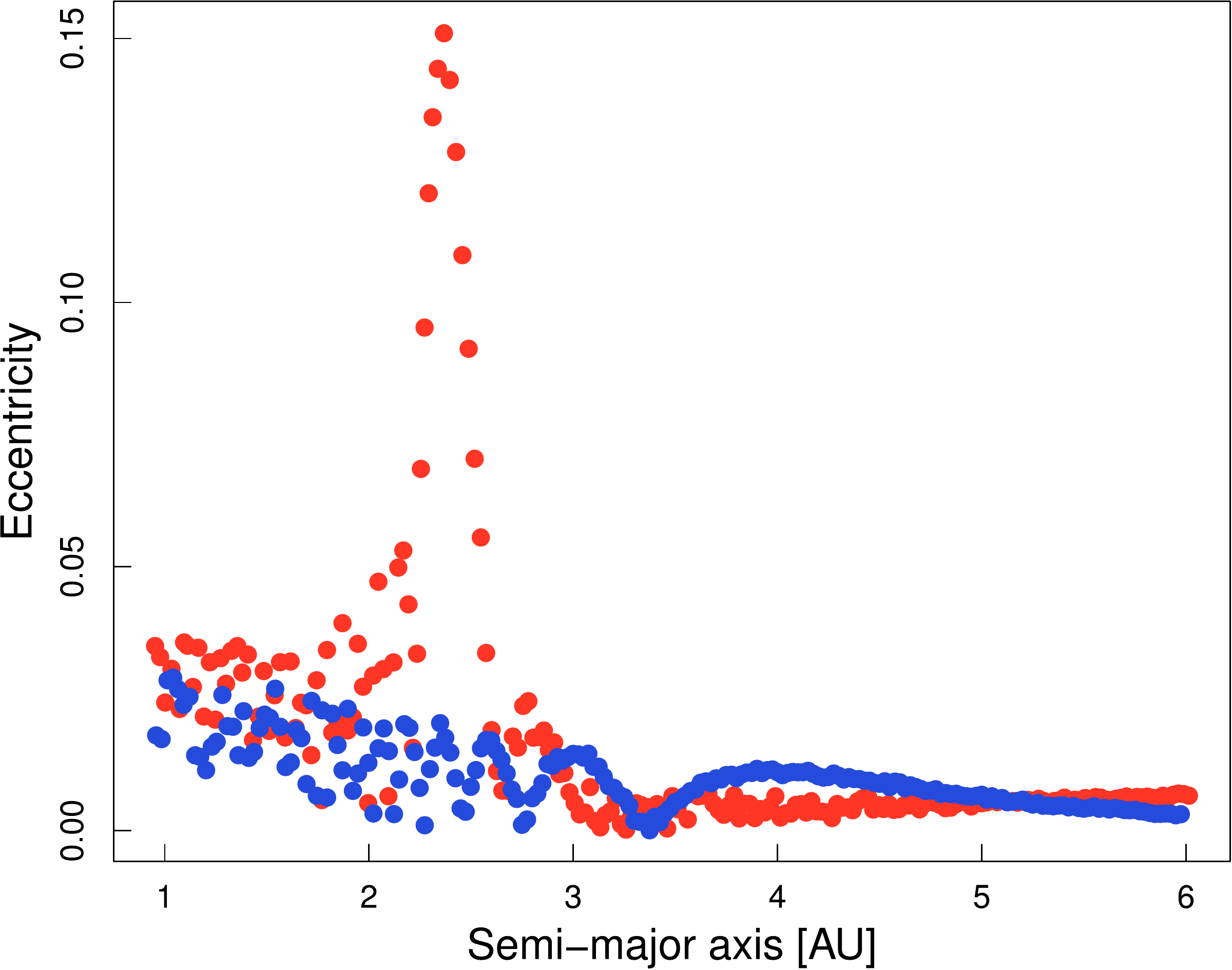}
\plotone{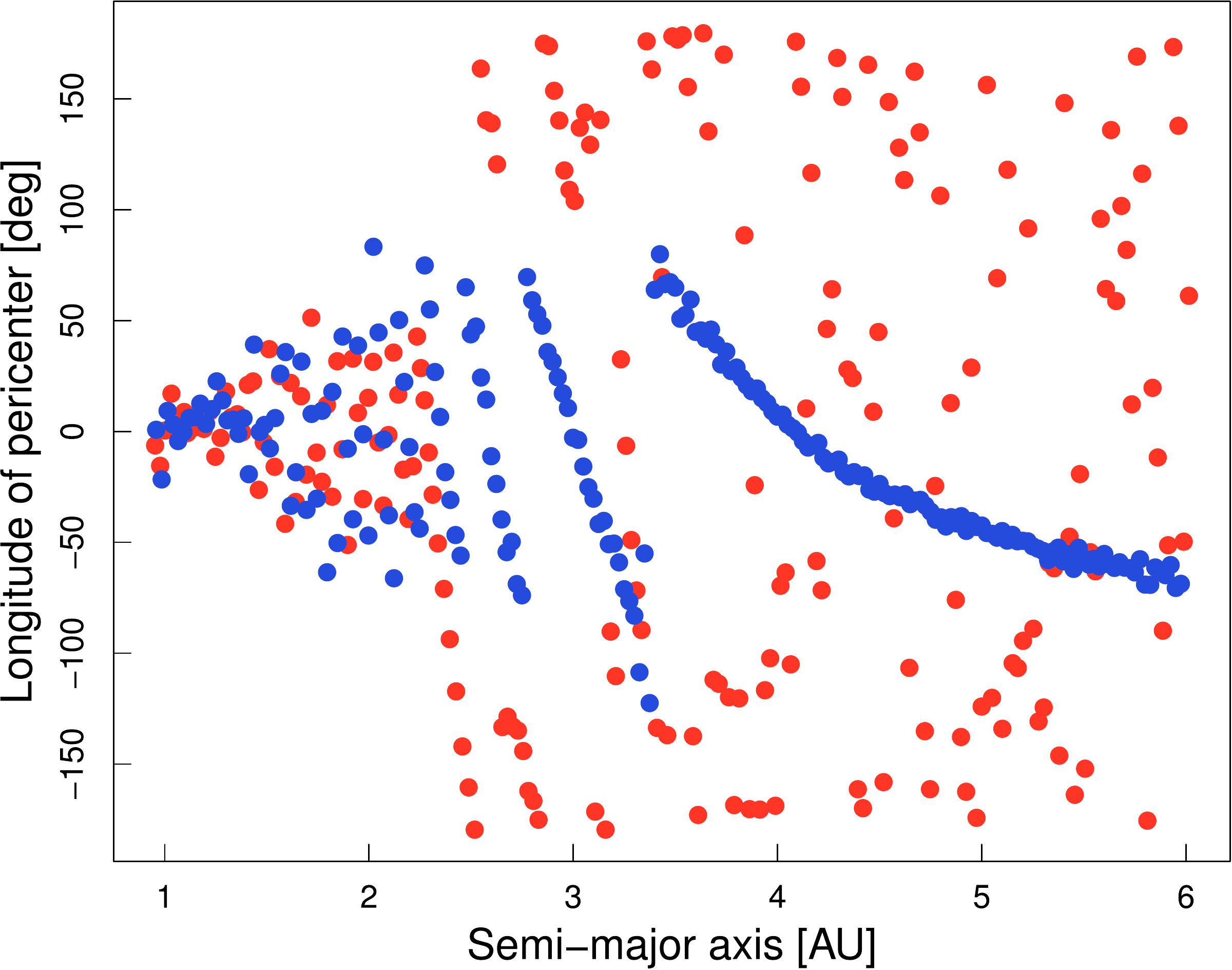}
\caption{Eccentricity and longitude of periastron after 10,000 years. Red points represent the model including the gas potential, while blue points represent the model neglecting the gas potential. For this figure, we consider 5-km planetesimals for direct comparison with \citet{Marzari13}.}\label{fig:ecc_lop}
\end{figure}

To investigate planetesimal evolution within a more realistic gas background, we subsequently ran a simplified model where the surface density profile is fixed to that derived from the hydrodynamical snapshot at $t = 10^4$ years (i.e. ignoring any further viscous evolution of the disk for the duration of the simulation). While this is approximation is unwarranted (given that significant viscous evolution is bound to happen over the time range explored below), it speeds up our code, since the gridding procedure can be run once at the beginning of the simulation (every hydrodynamical quantity is constant). To compute the hydrodynamical quantities precessed by $\pomega_\s{g}$, we simply rotate the azimuthal angle accordingly. 

For the fixed surface density model, we additionally take the gas disk potential $\Phi_\s{g}$ into account. \citet{Marzari13} showed that including the disk potential can raise the eccentricity of the planetesimals, potentially increasing the drift speed. The disk potential is calculated by direct summation once at the beginning of the simulation, and then rotated according to the precession frequency.

Figure \ref{fig:vr} shows the drift speed of 2.5-km planetesimals, with and without the gas potential. The drift speed is increased by a factor of 2 compared to the axisymmetric, static background. This increase in drift speed for eccentric, precessing disks was anticipated in the analytic models of \citet{Beauge10}. For the model including the gas potential, there is a spike in drift speed between approximately 2 and 3 AU; this is due to an enhanced planetesimal eccentricity in that region (see Figure \ref{fig:ecc_lop}). This region of eccentricity excitation is a secular resonance located where the secular precession amplitude $A$ equals the planetesimal precession $\dot\varpi_d$ induced by the gravity of the gas disk \citep[][note that we do not take the binary precession term $\dot\varpi_b$ into account]{Rafikov13}. Evaluating the  the secular equations of \citet{Rafikov13} with the parameters of our simulations yields a location for the resonance $r_{resonance} \approx 2.4$ AU, which is approximately consistent with Figure \ref{fig:ecc_lop}. The increase in eccentricity was also observed in \citet{Marzari13}. In this narrow region, planetesimals are rapidly depleted.

The drift speed decreases steeply (and actually changes sign) at $a_\s{stop} \approx 0.8$ AU ($\approx 3.5 a_\s{B}$), very close to the current observed location of the planet. This semi-major axis is close to the location of the pressure maximum ($R_\s{max} \approx 0.9$ AU). We expected that $a_\s{stop}$ and $R_\s{max}$ would not precisely coincide, since Equation \ref{eqn:gasspeed} is derived in an axisymmetric approximation (neglecting the eccentricity of the planetesimals and the gas, and the time-dependent potential of the central binary). We observe that planetesimals appear to smoothly halt at $a_\s{stop}$, and once they reached that radius, they show little semi-major axis variation (Figure \ref{fig:drift}, bottom panel). 

\subsection{Evolving surface density model}\label{sec:evolving}
\begin{figure}
\epsscale{\figscaletwo}
\plotone{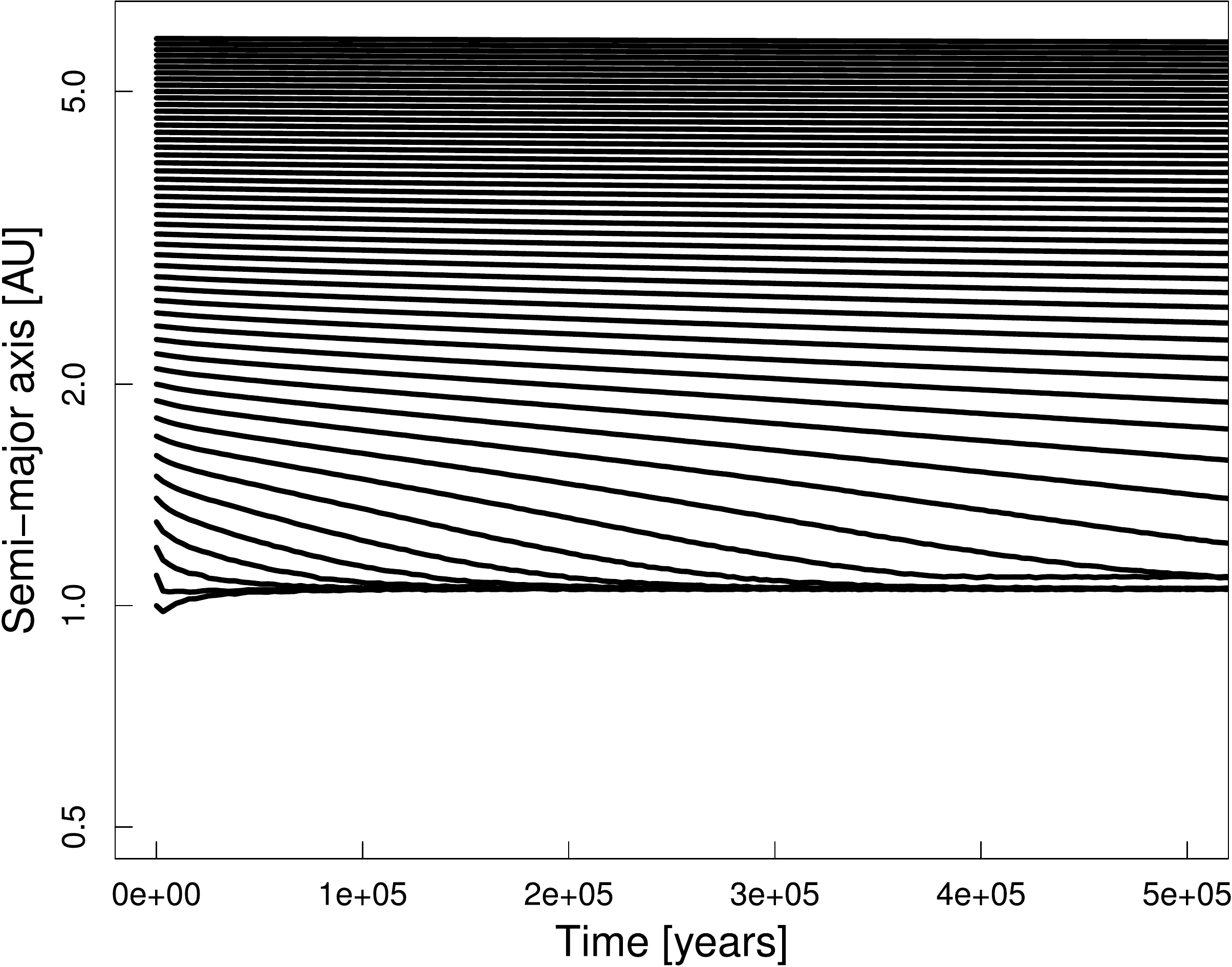}
\plotone{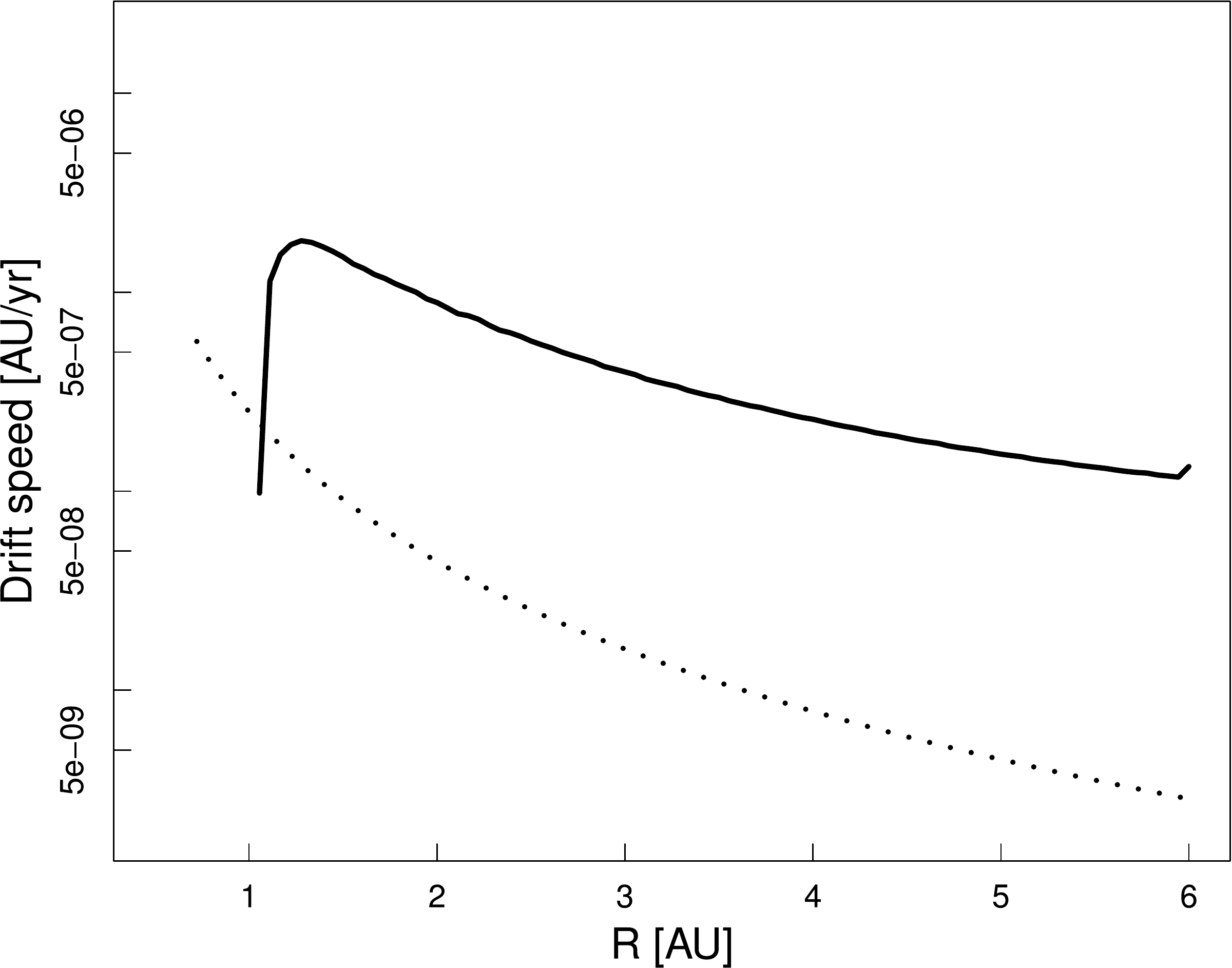}
\caption{\figp{Top} Planetesimal drift over $5\times 10^5$ years assuming the surface density evolution dictated by Equation \ref{eqn:onedev}. \figp{Bottom} Drift speed for $\radius_\s{pl} = 2.5$ km planetesimals. Also plotted is the drift speed in a  single-star environment of Figure \ref{fig:vr} for reference (dotted line, assuming zero eccentricity and a fixed surface density). }\label{fig:drift_ev}
\end{figure}

\begin{figure}
\epsscale{\figscalethree}
\plotone{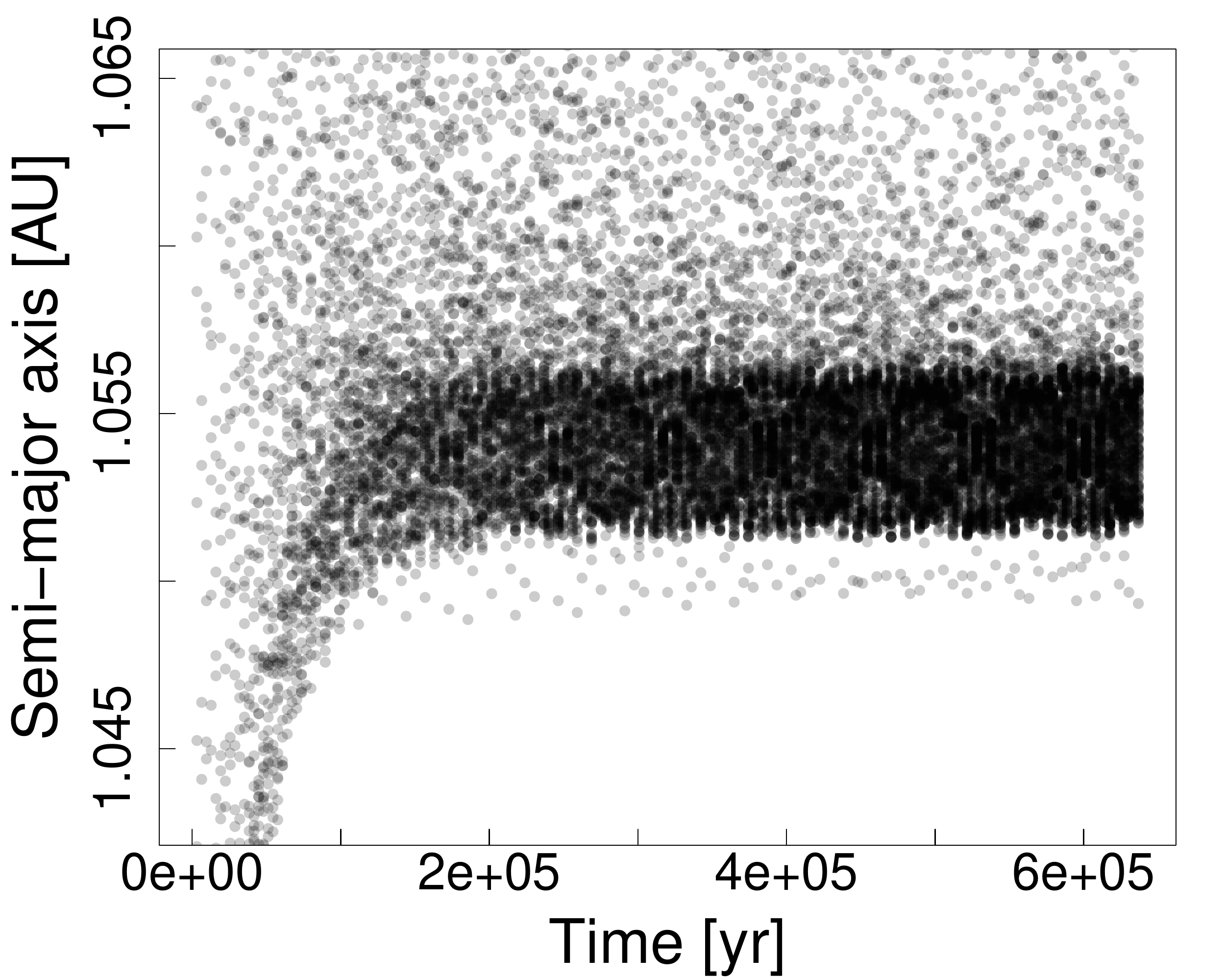}\\
\plotone{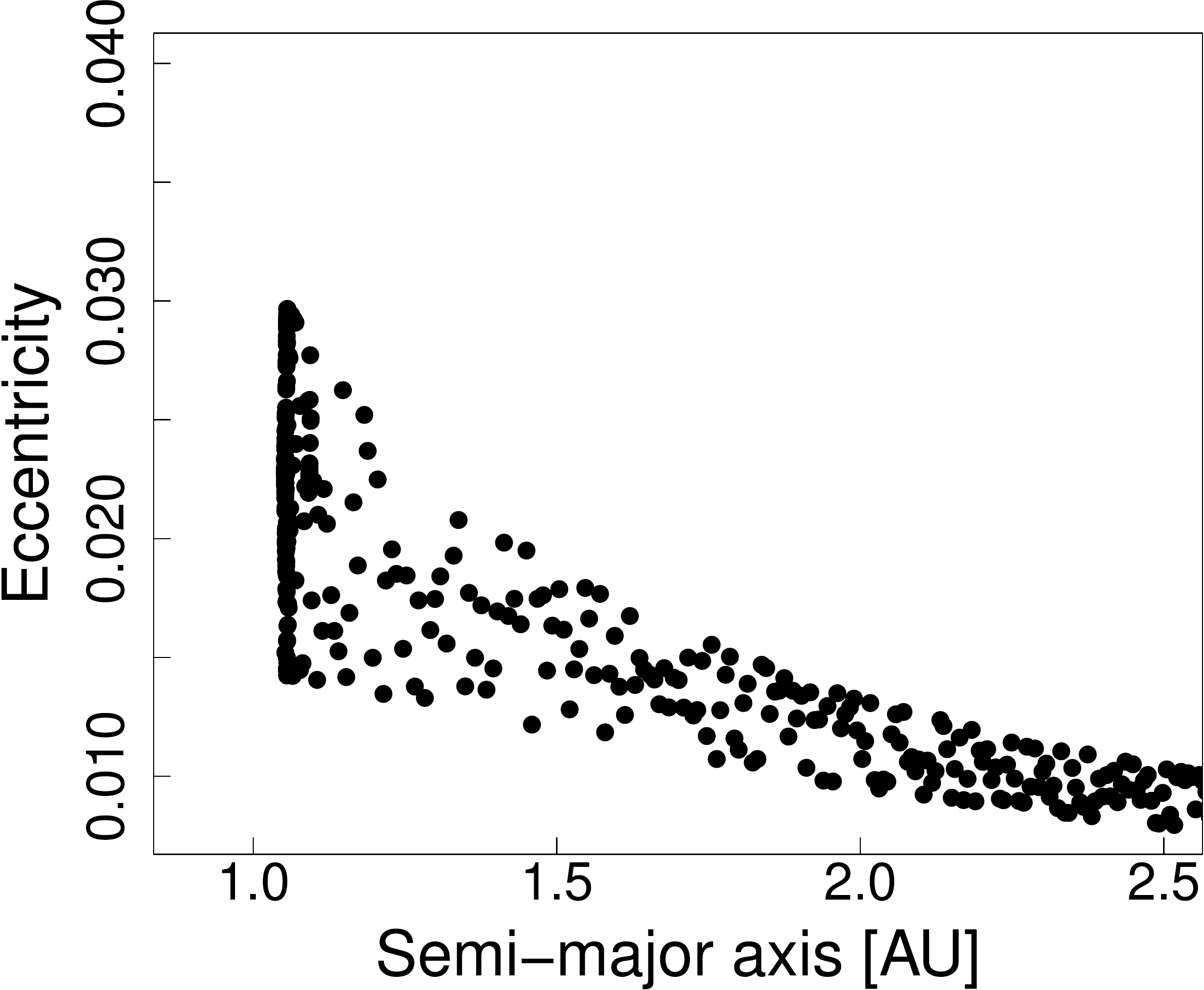}\\
\plotone{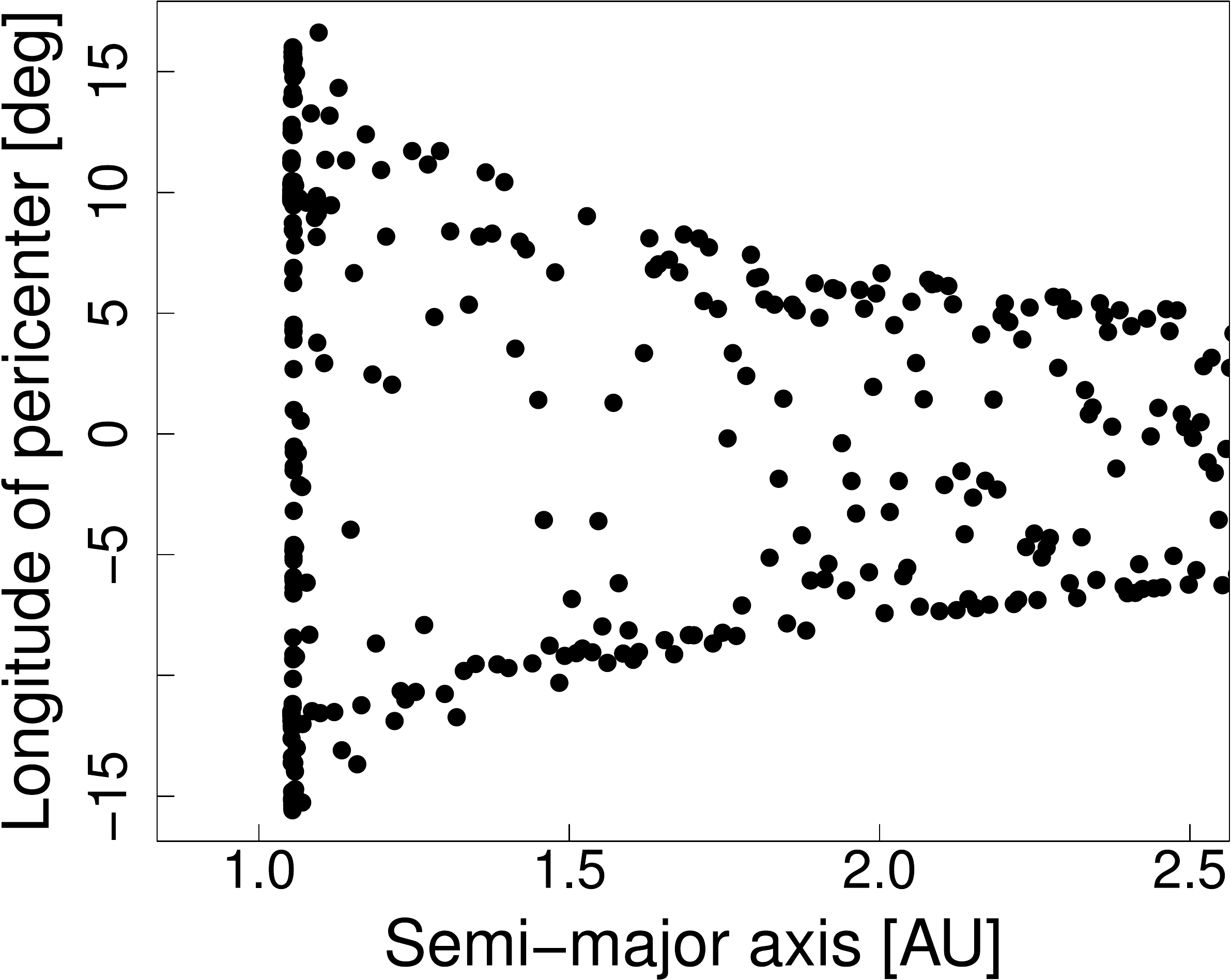}
\caption{\figp{Top} Accumulation of planetesimals in a confined belt around $a \approx 1$ AU. The semi-major axis of $\radius_\s{pl} = 2.5$ km planetesimals is plotted as a function of time. Note that planetesimals both inside and outside the belt converge quickly into the belt. \figp{Middle and bottom}
Eccentricity and longitude of periastron of 2.5-km planetesimals at $t = 5\times 10^5$ years.}\label{fig:tr_ecc_lop}
\end{figure}

As mentioned at the beginning of the previous subsection, ignoring the viscous evolution of the disk is an unwarranted simplification, since the viscous timescale $t_\s{disc} \sim a^2/\nu \approx 2\times 10^5$ years at 2 AU. Therefore, we need a recipe for evolving $\Sigma(\ve{r})$ as a function of time.

We model the surface density of the disk with the usual viscous evolution equations \citep{Lin86}, with the addition of an angular momentum source term representing the binary torque \citep{Pringle91, Armitage02, Alexander12}. Again, we effectively assume in our formulation that the surface density is approximately constant on each eccentric streamline (labeled by $a$), rather than on a circular radius $R$. Therefore, for the sake of the simplified model presented in this paper, we write the one-dimensional evolution equations as:
\begin{equation}
\frac{\partial\Sigma}{\partial t} = \frac{1}{a} \frac{\partial}{\partial a}\left[3a^{1/2}\ \frac{\partial\nu \Sigma a^{1/2}}{\partial a} - \frac{2\Lambda \Sigma a^{3/2}}{\left[\mass_1 + \mass_2\right]^{1/2}}\right] \ .\label{eqn:onedev}
\end{equation}
The term $\Lambda$ represents the torque exerted by the central binary and is written as \citep{Armitage02}:
\begin{equation}
\Lambda = \frac{b q^2 (\mass_1+\mass_2)}{2a}\left(\frac{a_\s{b}}{\max(H, |a-a_\s{b}|)}\right)^4\ ,\label{eqn:torque}
\end{equation}
where $q = \mass_2/\mass_1$ is the binary mass ratio. 

The surface density profile predicted by Equation \ref{eqn:onedev} generates surface density profiles that are more depleted at small radii ($a \lesssim 0.8$ AU) than the one derived from the hydrodynamical simulation at equal times. This is likely due to the fact that the torque prescribed by Equation \ref{eqn:torque} does not allow any accretion on the disk and quickly pushes material in the inner disk outwards, while in our full hydrodynamical simulations material continues to accrete onto the binary via non-axisymmetric streams. We set the dimensionless parameter $b\approx 0.3$ to match the density profiles generated by Equation \ref{eqn:onedev} with the outputs of our FARGO runs (Section \ref{sec:fargo}). While not a perfect match, it is adequate for the order-of-magnitude model presented in this paper.

The viscous evolution dictated by Equation \ref{eqn:onedev} quickly reduces the gas surface density during the first few $10^4$ years; correspondingly, drift speeds are also reduced.  We therefore consider a more massive disk \citep[2 times the standard MMSN normalization of][]{Hayashi81}, such that drift speeds are comparable to those observed in the previous section. 

We couple this 1D model to the \SPHIGA{} code, and update $\Sigma(\ve{r})$ (and, therefore, $v_\s{rel}$) at each time-step. We neglect the gas potential (which would also need to be recalculated at each time-step) to ease the computational burden. The previous section showed that the main consequence of including the gas potential was to speed up drift between 2 and 3 AU, while drift speed was essentially the same elsewhere in the disk. 

The planetesimal drift in this model is shown in Figure \ref{fig:drift_ev}. Drift speeds are comparable to those observed in the previous section. Planetesimals in the outer disk drift inwards and tend to converge to the trap at $\approx 1$ AU.  Planetesimals with initial semi-major axes inside the trap rapidly drift outwards, with a speed that increases sharply at small radii (due to the steep decrease in surface density, which results in a steep positive pressure gradient). 

The top panel of Figure \ref{fig:tr_ecc_lop} shows the distribution of semi-major axes of planetesimals throughout a simulation. As noted in the previous section, planetesimals that are captured in the belt tend to be confined to a narrow range in semi-major axis ($\Delta a \approx 0.005$ AU). We note that trapped planetesimals occupy a well-defined locus in eccentricity and longitude of pericenter as well. The bottom panel of Figure \ref{fig:tr_ecc_lop} shows a snapshot of the simulation ($t = 5\times 10^5$ years) where planetesimals in the belt are confined to $e \in [0.015, 0.03],\ \pomega - \pomega_B \in [-15^\circ, 15^\circ]$ (i.e. aligned with the central binary). 

\section{Planetesimal growth in the inner disk}\label{sec:results}
In Section \ref{sec:drift}, we verified that km-sized planetesimals in a circumbinary disk will tend to drift inwards, until they encounter a pressure maximum. At the pressure maximum, the aerodynamic drag vanishes; therefore, solids tend to be stranded in a well-confined ``belt'' (close to 1 AU for the nominal set of disk parameters).  Smaller debris (``dust'') produced by planetesimal-planetesimal destructive collisions will also tend to converge to the same region. Therefore, surviving planetesimals that drift into the belt will find themselves in a solid-rich region. Assuming that the efficiency of accreting dust is high, planetesimals could conceivably accrete mass from the debris reservoir to become indestructible ($\radius_\s{pl} \approx 100$ km) and subsequently form the building blocks of a planetary core. The crux is now establishing that this process is quick enough to produce a core within a reasonable time scale.

\subsection{Numerical model}
A rigorous approach to the problem at hand would involve tracking planetesimal destruction and dust accretion self-consistently in the manner of \citet{Paardekooper08} and P12. In their simulations, the cascade of fragments produced by collisions is followed down to a certain size, where it is deemed as ``dust''. 

The numerical model of P12 has a number of downsides. Firstly, it is assumed that there is no pressure gradient in the static background disk (an ad-hoc assumption to simplify the model); therefore, in their simulations planetesimals and dust do \textit{not} drift. Secondly, dust produced by the planetesimal collisions is accumulated in-place into circular radial bins. Rather, we should expect dust (which is strongly coupled to the gas through aerodynamic drag) to settle into the eccentric streamlines followed by the gas disk. Finally, the computational burden of tracking collisions and accretion within the $N$-body code of Section \ref{sec:drift} becomes rapidly unmanageable as the number of fragments increases. 

The problem at hand does offer a number of simplifications, however. Collisional rates will be high throughout the disk due to the high encounter speeds observed both in our models and the simulations of \citet{Marzari13}, and the collisional outcome will always be destructive. Planetesimals will be ground into many sub-km fragments by collisions rather quickly; for example, the largest fragment produced from the collision of two primordial $\radius_\s{pl} = 5$ km planetesimals at $\approx 200$ \ms{} (super-catastrophic regime) is only $\approx 800$ m in radius. Subsequent collisions will further grind the material. If we assume a realistic pressure profile (as opposed to the no-drift condition of P12), then the sub-km fragments should be removed from their original site and spiral into the pressure maximum on a very short timescale.  Therefore, if we assume that planetesimals are rapidly ground into small debris which drift into the belt on a short timescale, then we can crudely assume that planetesimal collisions result in an ``instantaneous'' increase of dust mass in the belt. This absolves the code from having to track destruction and accretion throughout the disk,  representing the main simplification of our model.

We now discuss a simple back-of-the-envelope model that will attempt to qualitatively represent the evolution of the planetesimal system in presence of collisions, drift and dust accretion (as sketched in the Introduction). It couples a single PDE following planetesimal drift and destruction outside the belt with a stochastic model for planetesimal size evolution inside the belt. This model trades some fidelity for execution speed, in order to run the simulation for several $10^5$ years; we discuss some of its limitations in Section \ref{sec:caveats}. 

Below, we explain the ingredients of the model in more detail. The bulk of our model are informed by the $N$-body simulations presented in the previous Section. However, a few parameters are not constrained by our simulation; we list these parameters and their nominal values in Table \ref{tab:pars}.

\begin{deluxetable}{lllllll}
\tablehead{\colhead{}& \colhead{Meaning} & \colhead{Simulation values}}
\tablecaption{}
\startdata
$\rho_\s{pl}$ 		& Planetesimal density & 2 g cm$^{-3}$	\\
$h$				& Scale height normalization & 0.05 \\
$\radius_\s{pl, 0}$	& Initial planetesimal radius	& {5} km\\
$\Sigma_\s{pl, 0}$	& Density of solids at 1 AU	& 5, \textbf{10}, 20 g cm$^{-2}$\\
$\beta_\s{1}$		& Vertical thickness of dust annulus $\dagger$	& {1}, \textbf{1/15}, 1/150 \\
$\beta_\s{2}$		& Radial extent of dust annulus	$\dagger$ & \textbf{1}, 2,  5, 10 \\
$\epsilon$		& Dust accretion efficiency	& {1}\\
\enddata
\tablecomments{
Nominal values are bolded.
$\dagger$ Normalized by the local gas scale height $H$}
\label{tab:pars}
\end{deluxetable}

\subsection{Planetesimal destruction and drift}
We divide the planetesimal disk in two regions: the semi-major axis range $R_\s{b} \pm \Delta R_\s{b}$ (the belt, centered on $R_\s{b}$ and with thickness $\Delta R_\s{b}$) and the rest of the disk.  We assume that a single-sized population of planetesimals with radius $\radius_\s{pl, 0}$ (mass $\mass_\s{pl, 0}$) are distributed throughout the disk with a number surface density $\nsd_\s{pl}$. As seen in Section \ref{sec:drift}, the planetesimal disk will be endowed with a small eccentricity. For simplicity, however, we will assume that the surface density of planetesimals is axisymmetric ($\nsd_\s{pl} \equiv \nsd_\s{pl}(R)$). The planetesimal eccentricity factors into the planetesimal collision speed $v_\s{coll}(R, \radius_\s{pl, 0})$ and the planetesimal drift speed $v_\s{drift}(R, \radius_\s{pl, 0})$, which are estimated from the simulations conducted in Section \ref{sec:evolving}.

We write the time evolution of the surface density as
\begin{equation}
\frac{\partial \nsd_\s{pl}}{\partial t} = -\frac{\partial \nsd_\s{pl}}{\partial t}\vert_\s{coll} - \frac{1}{R}\frac{\partial}{\partial R} R v_\s{drift}\nsd_\s{pl}\ \ .\label{eqn:pde_sd}
\end{equation}
This equation assumes that the time evolution is driven exclusively by destructive collisions (which reduces the surface density in planetesimals) and drift; in particular, planetesimals outside the belt do not grow. The collisional term $\partial \nsd_\s{pl}/\partial t \vert_\s{coll}$ can be written as
\begin{equation}
\frac{\partial\nsd_\s{pl}}{\partial t}\vert_\s{coll} = \frac{\nsd_\s{pl}^2}{2\bar{i}R} \pi \radius_\s{pl}^2 v_\s{coll} \ ,
\end{equation}
The main unknown parameter is the average inclination $\bar{i}(R)$ (i.e. the thickness of the planetesimal disk); we assume it is set by the escape velocity of planetesimals, $\bar{i}(R) \approx [(2/3) \pi R \rho_\s{pl} \radius_\s{pl}^2]^{1/2}$. The planetesimal-planetesimal collision speed $v_\s{coll}$ is instead measured from the $N$-body simulations: we evolve a disk of 5-km planetesimals for $2\times 10^4$ years (as in Figure \ref{fig:tr_ecc_lop}), and calculate the median collision velocity as a function of distance from the central binary, giving us an approximate $v_{coll}(R)$. 

The debris generated by the collisional term is assumed to rapidly drift into the belt (on a much shorter timescale than the planetesimal drift), adding into the dust surface density (Equation \ref{eqn:dust}). The solid surface density normalization determines the initial planetesimal surface density; we chose a nominal MMSN value of $\Sigma_{pl, 0} = 10$ g cm$^{-2}$.

Equation \ref{eqn:pde_sd} is solved with a straightforward finite difference method \citep[e.g.][]{Press92}. We take an initial distribution of planetesimals $\nsd_\s{pl} \propto R^{-1}$ extending between $R_\s{b}$ ($\approx 1$ AU) and 10 AU. The normalization of $\nsd_\s{pl}$ is determined by setting the mass of solids present in the disk.

\subsection{Planetesimals in the belt}\label{sec:dustaccr}
At each time-step, a number of planetesimals will avoid destruction and drift into the belt (delimited spatially by $R_\s{b} \pm \Delta R_\s{b}$). We model this by calculating the planetesimal flux as dictated by Equation \ref{eqn:pde_sd} in a ghost zone overlapping the belt.

Planetesimals that drift in the belt are tracked as a set of $N$ evolving radii $\radius_\s{i} = {\radius_1, \radius_2, ..., \radius_\s{N}}$, i.e. we start following the size evolution of planetesimals only once they enter the belt. Dust accretion is calculated as
\begin{equation}
\dot{\radius_i} = \frac{\Sigma_\s{d}}{4\beta_\s{1} H \rho_\s{pl}} \Delta v_\s{d} \bar\epsilon\ ,
\label{eqn:radev}
\end{equation} 
i.e. assuming that each planetesimal ``sweeps'' through a dusty annulus with surface density $\Sigma_\s{d}$ and scale height $H_\s{d} = \beta_\s{1} H$ (a fraction $\beta_\s{1}$ of the local gas scale height $H$). The parameter $\beta_\s{1}$ parametrizes the degree of turbulence in the disk; a small $\beta_\s{1}$ indicates that the dust annulus has settled close to the midplane (thereby increasing the dust accretion rate), implying a low level of turbulence. We do not consider any enhancement in the accretion rate by gravitational focusing, since we assume that the dust is well coupled to the gas.

The efficiency of dust accretion is parametrized by $\bar\epsilon$, varying between 0 (no dust accretion) and 1 (perfect dust accretion).  We follow P12 in assuming perfect dust accretion whenever the relative velocity between the dust ring and planetesimals is larger than 100 times the planetesimal escape velocity. Since the dust ring (which follows the eccentric gas streamlines) will not, in general, be aligned with the planetesimal orbit, we average $\epsilon$ and $\Delta v_d$ along the orbit of planetesimals and over the ensemble of planetesimal orbital parameters (taking $\epsilon = 0$ when $\Delta v_\s{d} > 100 v_\s{esc}$ and $\epsilon = 1$ otherwise). The resulting accretion efficiency depends on the planetesimal radius; as the planetesimal increases in size, the efficiency will become larger until it reaches $\bar\epsilon = 1$ (at $\radius_\s{pl} \approx 30$ km).

Finally, we need to take into account planetesimal-planetesimal collisions within the belt. We calculate the probability for each planetesimal $i$ tracked that it will experience one or more collisions during a time-step $\Delta t$ as:
\begin{equation}
\prob_i = 
\frac{\nsd_\s{pl}}{2 \bar{i} R} \pi \radius_\s{pl}^2 \Delta v_\s{coll} (1+F) \Delta t\ ,
\end{equation}
where $\nsd_\s{pl} \approx N/A_\s{b}$ ($A_\s{b} = 2\pi R_\s{b} \Delta R_\s{b}$ being the area covered by the belt), $\Delta v_\s{coll}$ is the collision speed, and $F = (8\pi/3)  \radius_\s{i}^2 \rho_\s{pl} \Delta v_\s{coll}^{-2}$ is the gravitational focusing factor. We subsequently draw a uniform random number $u$, and a collision is experienced if $u < \prob_i$ (we ensure $\max_i(\prob_i) \ll 1$ by reducing the time-step as needed). $\Delta v_\s{coll}$ is drawn from the range of collision speeds between 5-km planetesimals trapped in the belt, as measured by the $N$-body simulations. In a strict sense, as planetesimal seeds grow to larger sizes, the distribution of $\Delta v_\s{coll}$ should also evolve; however, we stop the simulation before planetesimals grow large enough to significantly perturb the background planetesimal population (see also Section \ref{sec:results_strongweak}). 

If a collision is experienced, then a second planetesimal $j$ is selected randomly. When $\Delta v_\s{coll} < v_\s{esc}$, we add the mass of the second planetesimal to the first and remove the second planetesimal from the simulation. When $\Delta v_\s{coll} > v_\s{esc}$, the outcome of a collision is decided using the disruption criteria of \citet{Stewart09} and \citet{Leinhardt12}. This recipe gives us the size of the largest remaining fragment $\radius'$ as a function of $\radius_\s{i}$, $\radius_\s{j}$ and $\Delta v_\s{coll}$. We consider two regimes: ``strong'' planetesimals and ``weak'' planetesimals, each possessing different material strengths \citep{Thebault11}. 

\subsection{Dust generation}
The last ingredient of our model is the accumulation of dust in an annulus overlapping the planetesimal belt. We assume that debris produced by planetesimal grinding (throughout the disk and in the belt) will be small and therefore quickly siphoned into a uniform annulus overlapping the planetesimal belt. We expect the dust to be will be well-coupled to the gas and follow the gas streamlines once settled close to the pressure maximum.  

The surface density of the dust annulus is given by

\begin{equation}
\dot{\Sigma}_\s{d} = \frac{2\pi}{A_\s{d}} \left[ \int R dR \frac{\partial A_\s{pl}}{\partial t}\vert_\s{coll} \mass_\s{pl} - \sum_\s{i} \frac{2}{3} \radius_\s{i}^2 \dot{\radius}_\s{i} \rho_\s{pl}\right] - \frac{\Sigma_\s{d}}{t_\s{d}}  \ ,
\label{eqn:dust}
\end{equation}
where the first term represents dust generation by planetesimal grinding, and the second term is dust accretion by planetesimals in the belt (Equation \ref{eqn:radev}). $A_\s{d}$ is the area of the dust annulus. The width of the dust annulus $\Delta r_\s{d}$ cannot be determined from our simulations. It will likely be set by the competition between the small-scale interaction with the gas disk (i.e. turbulent motions in the radial direction, and Brownian motions due to collisions with the gas molecules), which tends to smear the annulus, and the overall pressure gradient, which tends to concentrate the dust towards the pressure maximum. We take the relevant radial scale to be set by $H$ (the scale of the largest turbulent eddies), so that $\Delta r_\s{d} = \beta_\s{2} H$, with a nominal value $\beta_\s{2}=1$; we also consider higher values of $\beta_\s{2}$, which are less favorable to dust accretion. Finally, the third term takes into account the possible loss of dust. The binary torque will tend to stop inflow of material; however, some accretion will still take place through non-axisymmetric streams. The accretion rate in the simulations of \citet{MacFadyen08} is approximately 10\% of the steady flow accretion rate. For lack of better guidance, we assume that the dust accretion timescale $t_\s{d} = 10 t_\s{\nu}$, where $t_\s{\nu}$ is the local viscous timescale.

One last source of uncertainty is the extent of the radial migration of the dust. Dust produced in the disk outside the snow line ($\approx 3$ AU) could potentially be trapped at the snow line \citep[e.g.][]{Kretke07}. Modeling the detailed disk structure is beyond the scope of this paper; therefore, we cautiously only consider the inner 3 AU of the disk as sources of dust that migrates into the annulus at 1 AU. This limitation sets a maximum mass that can participate in the formation of the core.

\begin{figure}
\epsscale{\figscalethree}
\plotone{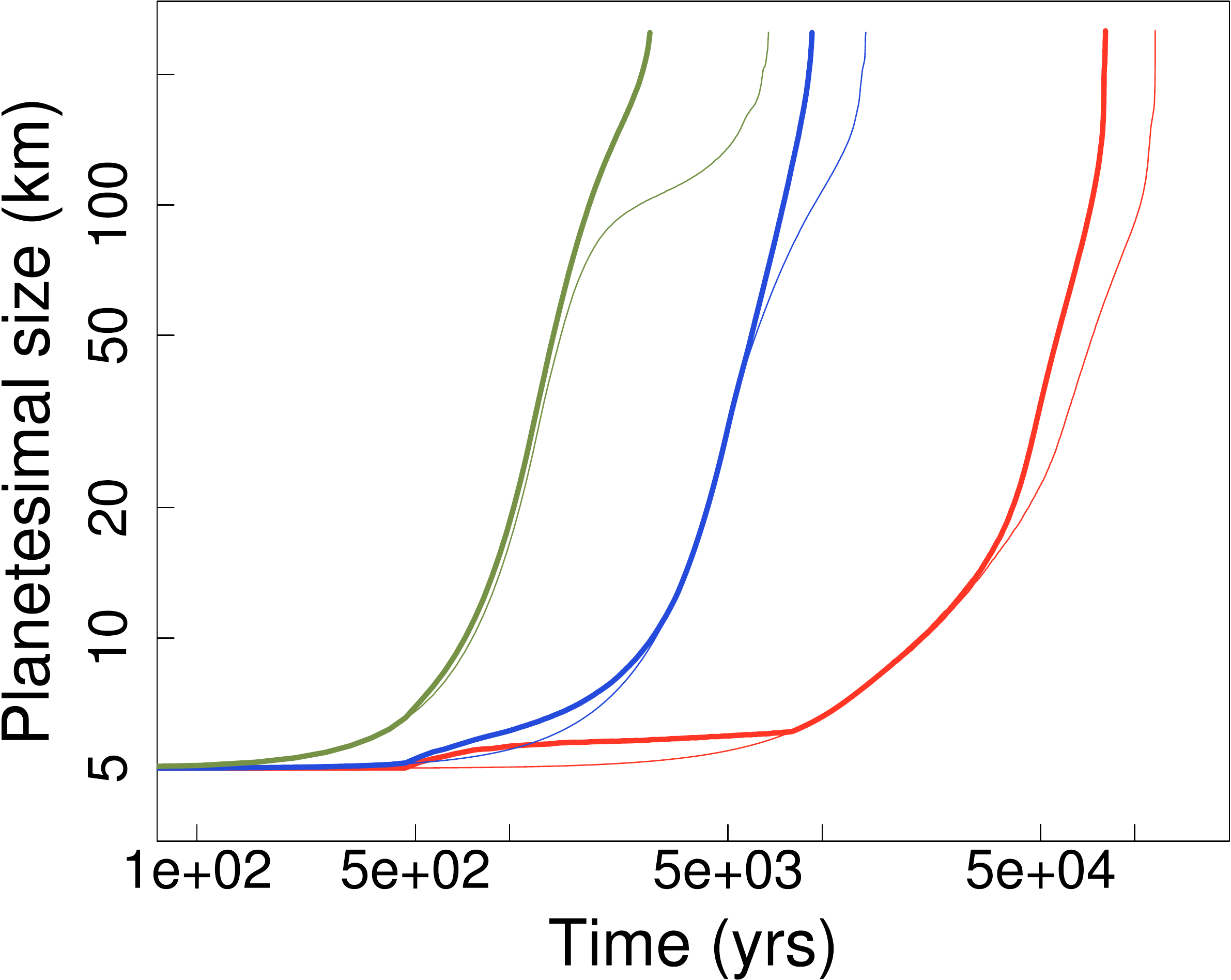}\\
\plotone{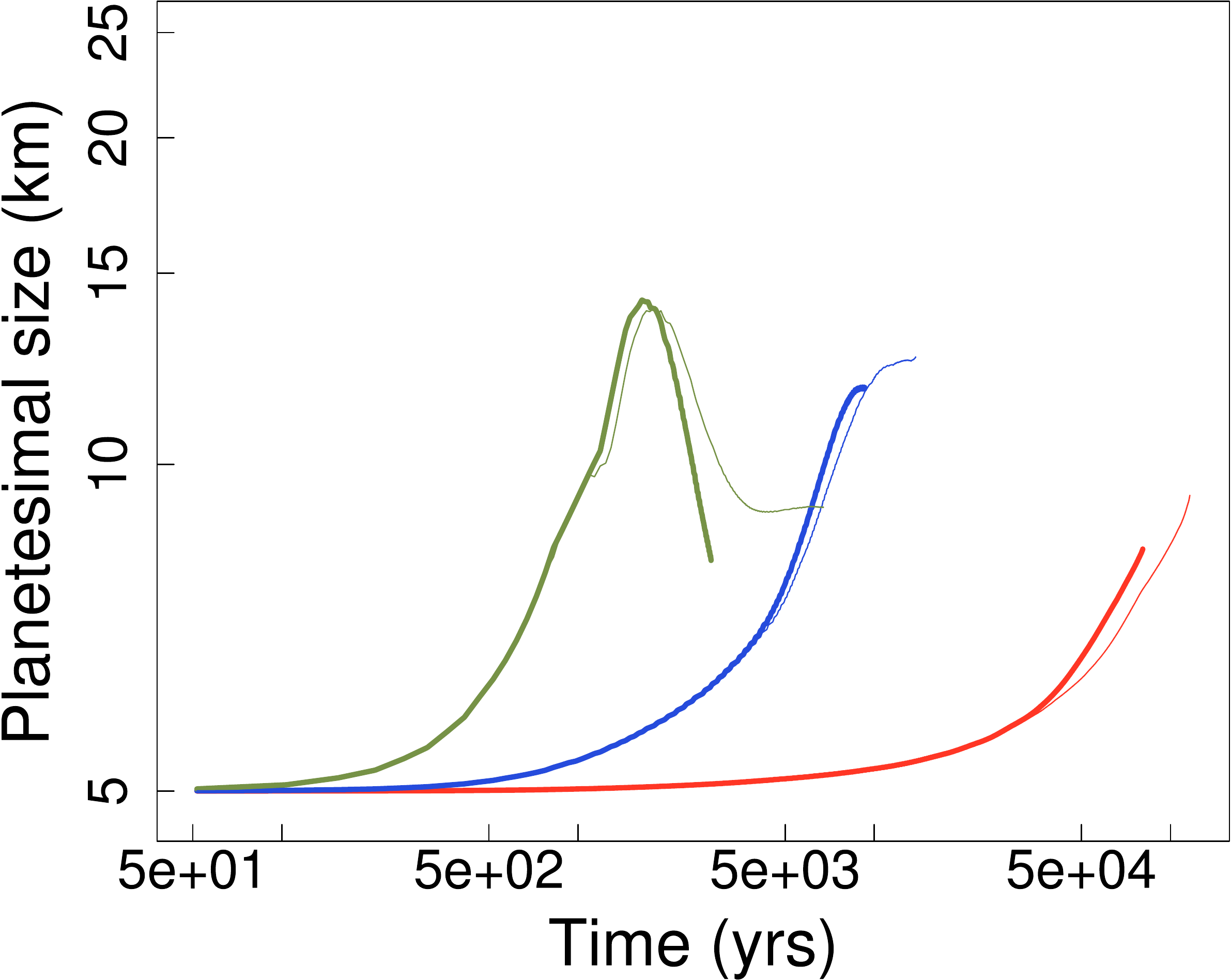}\\
\plotone{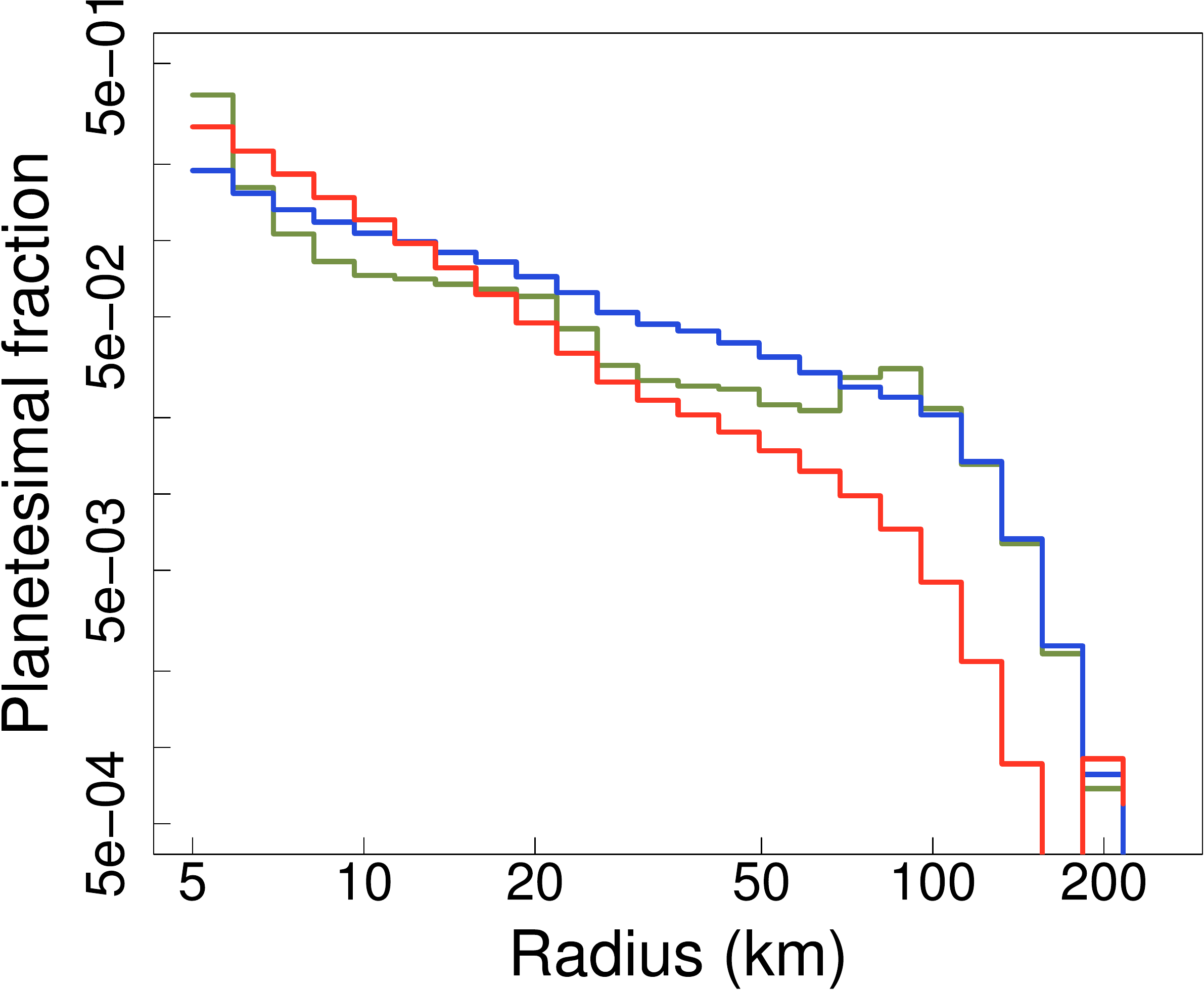}
\caption{\figp{Top} Radius of the 1000-th largest body as a function of time, in presence of strong (red), intermediate (blue) and weak (green) turbulence. The corresponding size evolution for weak aggregates is plotted with a thin line. \figp{Middle} Median radius of the planetesimal population as a function of time. \figp{Bottom} Size distribution of strong planetesimals in the belt at the end of the simulation.}\label{fig:rpl}
\end{figure}

\subsection{Results: strong and weak planetesimals}\label{sec:results_strongweak}
We first consider two limiting cases for the material parameters, representing ``strong'' planetesimals and ``weak'' aggregates \citep{Stewart09}. The choice of material parameters dictates the characteristic size above which collisions are not destructive (i.e. the mass of the largest fragment is larger than either of the planetesimals involved in the collision). For the median collision speed in the belt ($\approx 300$ m/s), a body immersed in a sea of 5-km bodies becomes indestructible when it reaches a size of $\approx$ 50 km ($\approx 200$ km for weak aggregates). Even as the largest bodies start becoming indestructible, two-body gravitational focusing remains weak due to the large velocity dispersion in the annulus ($v_\s{esc} \ll \Delta v_\s{coll}$). Runaway growth becomes significant when the largest bodies reach the critical size of  $\approx 250$ km (such that $F \approx 1$). We stop the simulation when there are at least 1000 bodies that have entered runaway growth. This critical size was chosen since once the largest bodies reach that size, they will start perturbing the background planetesimal population and increasing their velocity dispersion (and therefore increasing $\Delta v_\s{coll}$); therefore, we can no longer follow their size evolution using our simple model. We now describe the size evolution of our planetesimal population.

The top panel of Figure \ref{fig:rpl} plots the the size of the 1000-th largest body as a function of time (meaning that at any given time, there are 1000 ``seeds'' larger in radius).  We take the nominal value of $\Sigma_\s{pl, 0} = 10$ g cm$^{-2}$ (the solid surface density normalization) and consider three different values for $\beta_\s{1}$ (the thickness of the dust disk relative to the gas disk, a proxy for the amount of turbulence in the disk): strong ($\beta_\s{1} = 1$), intermediate ($\beta_\s{1} = 1/15$) and weak ($\beta_\s{1} = 1/150$) turbulence, following \citet{Xie10}. The strong turbulence value is the least favorable to planetesimal growth, since it limits the rate at which planetesimals can accrete dust. We plot the size evolution for strong and weak planetesimals; we expect the size evolution of a realistic planetesimal population to reside between the two extremes.

Initially, planetesimals in the belt grow exclusively by accreting dust. Dust accretion in our model is relatively fast, since the dust annulus is fed by planetesimal grinding throughout the disk, leading to high dust surface densities.  Still, the median size of planetesimals grows slowly (middle panel of Figure \ref{fig:rpl}). This is the result of the competition between dust accretion (which increases the radius only linearly; Equation \ref{eqn:radev}), the continual grinding by mutual collisions and the drifting of new 5-km planetesimals drifting in from the outer disk. 
On the other hand, a fraction of ``lucky'' planetesimals that suffer fewer collisions can continue growing to larger sizes, creating a size spectrum in the belt. Bodies residing in the tail of the size spectrum (the seeds) become large enough that a significant fraction of the possible impact speeds results in accretion rather than destruction. This opens a new growth channel that is not available to the sea of small planetesimals in the belt. Therefore, the indestructible seeds start growing at a much faster rate than the background planetesimals, and can very rapidly reach the runaway growth stage.

The timescale for reaching the critical size is determined primarily by $\beta_\s{1}$, spanning from just $\approx 2\times 10^3$ years ($\beta_\s{1} = 1/150$) to $\approx 10^5$ years ($\beta_\s{1} = 1$). This is due to the fact that $\beta_\s{1}$ sets the dust accretion rate, and consequently the time it takes for bodies to reach the indestructible size. The material strength also plays a role: strong seeds are able to start accreting other planetesimals at a smaller size than their weak counterparts, reaching the critical size earlier.

Finally, the bottom panel of Figure \ref{fig:rpl} shows a snapshot of the size distribution of planetesimals at the end of each simulation. We note that, again, the size distribution is primarily set by $\beta_\s{1}$: faster accretion rates imply that planetesimals are able to leave the 5-km size bin rapidly and become less susceptible to destructive encounters. Consequently, the size distribution becomes steeper as the turbulence parameter $\beta_\s{1}$ is increased. 

\begin{figure}
\epsscale{\figscaletwo}
\plotone{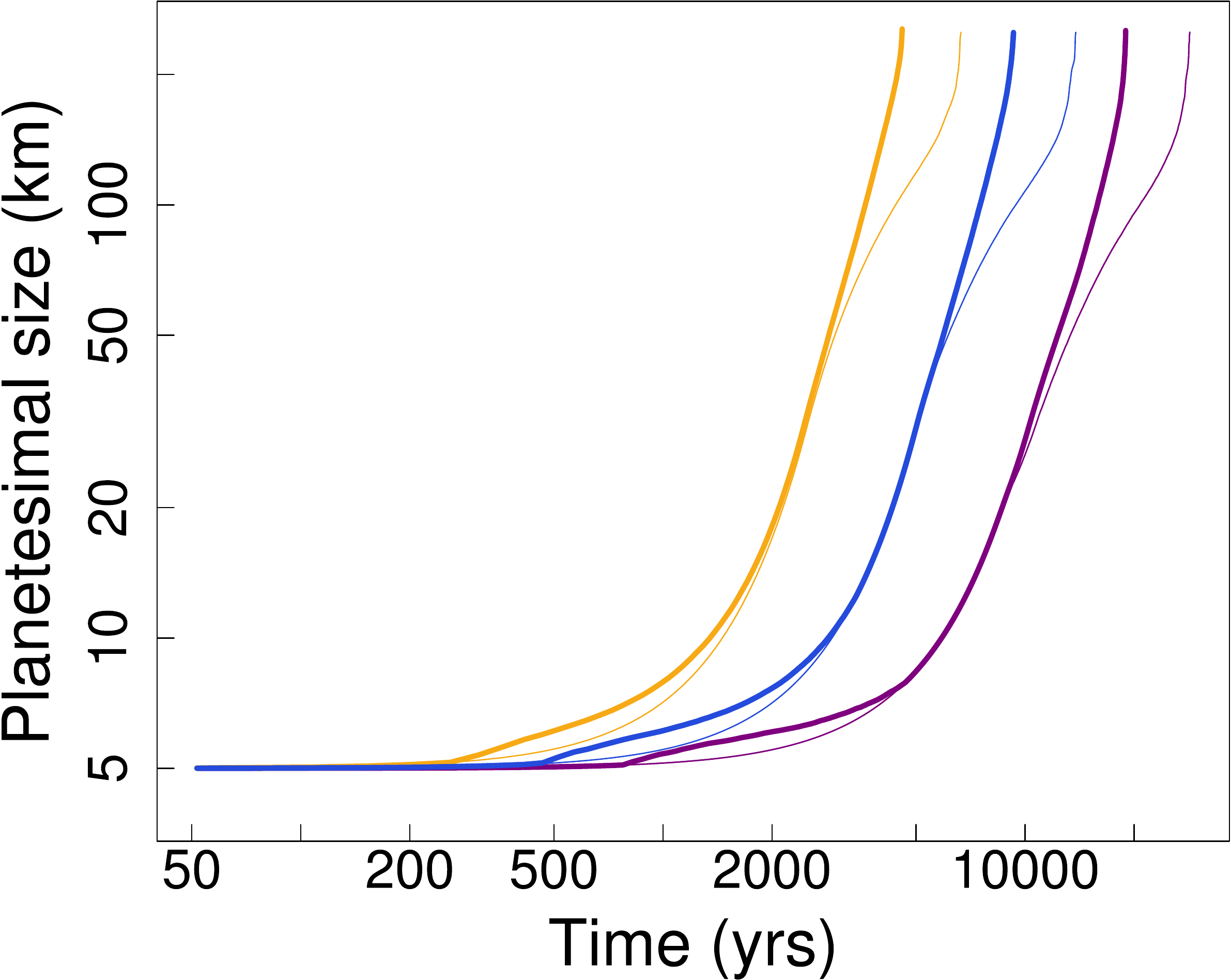}\\
\plotone{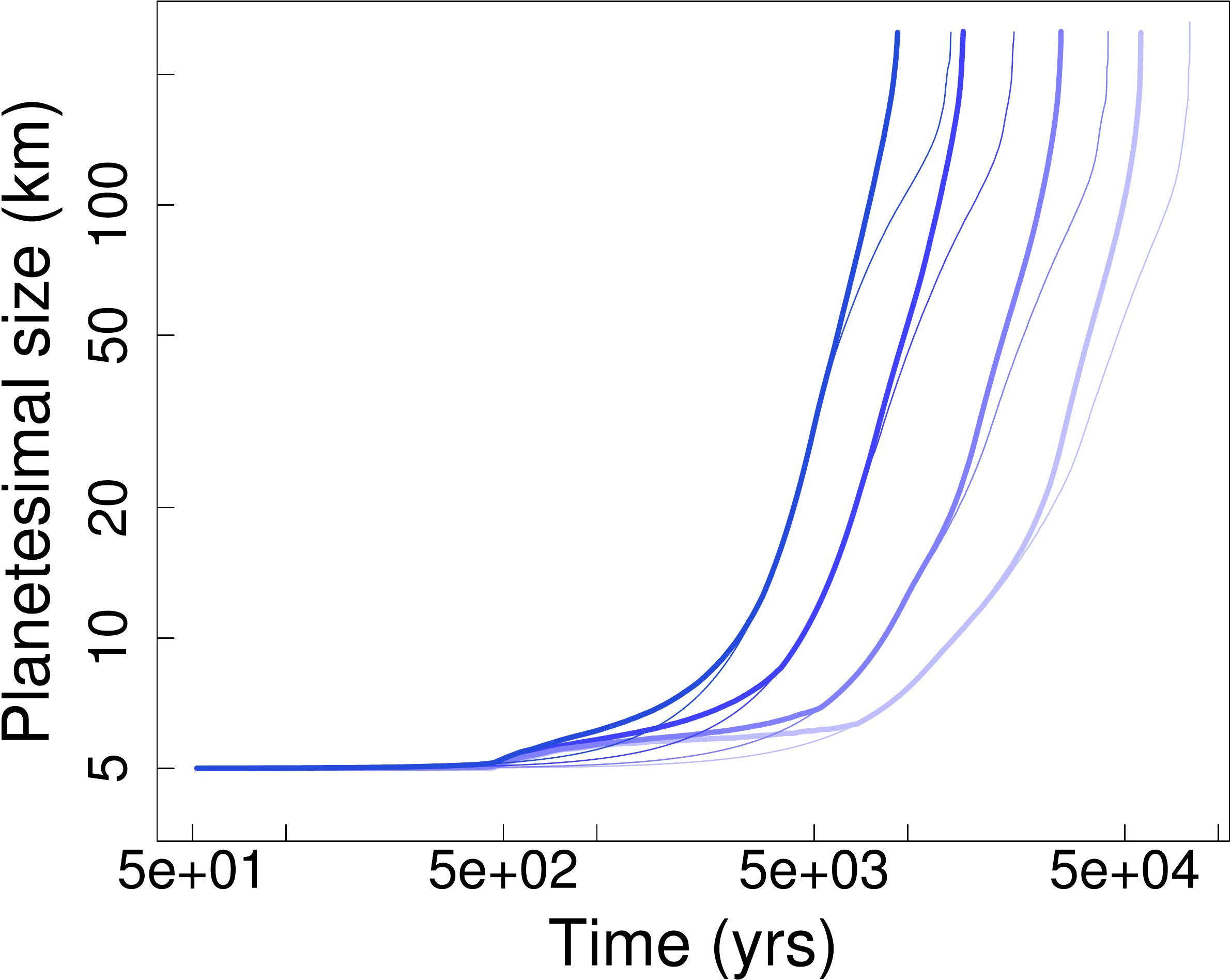}
\caption{(\textit{Top}) Radius of the 1000-th largest body as a function of time in presence of intermediate turbulence ($\beta_\s{1} = 1/15$), for $\Sigma_\s{pl, 0} = 5, 10, 20$ g cm$^{-2}$ (purple, blue and orange, respectively). (\textit{Bottom}) Radius of the 1000-th largest body as a function of time in presence of intermediate turbulence ($\beta_\s{1} = 1/15$) for $\beta_\s{2} = 1, 2, 5, 10$ (from left to right). $\beta_\s{2}$ sets the radial extent of the dust annulus in units of the gas scale height $H$. The corresponding size evolution for weak aggregates is plotted with a thin line.}\label{fig:rplsigma}
\end{figure}

We also considered three different normalizations for the initial solid surface density (5, 10, and 20 g cm$^{-2}$), where 10 g cm$^{-2}$ is our nominal MMSN value. The larger normalization reflects the likely possibility that the primordial solid inventory in an MMSN-like nebula was larger (by a factor of 2-3), but lost due to inefficiencies in the planet formation process \citep[e.g.][]{Hansen12}. On the other hand, the smaller normalization takes into account the possibility of a less solid-rich nebula; this normalization could be appropriate for the sub-solar metallicity of Kepler-16 ([Fe/H $= 0.3 \pm 0.2$). We plot the growth curves in the top panel of Figure \ref{fig:rplsigma}, where we took $\beta_\s{1} = 1/15$ (corresponding to an intermediate turbulence level). As expected, a larger initial surface density results in a faster growth, due both to the higher surface density of the dust annulus and the larger number of planetesimals available to accrete. Similarly, in the bottom panel of Figure \ref{fig:rplsigma} we considered three different normalizations for the radial extent of the dust annulus $\Delta r_d = \beta_\s{2} H$; higher values of $\beta_\s{2}$ reduce the surface density of dust $\Sigma_d$, and therefore slow down the initial accretion of dust by the planetesimals. 

We note that the dust accretion rate determines the timescale for the onset of planetesimal accretion (and the subsequent runaway phase). The dust accretion rate depends on the ratio of $\Sigma_d$ and $\beta_\s{1}$ (Equation \ref{eqn:radev}), and ${\Sigma}_\s{d}$ itself depends on $\beta_\s{2}$ (Equation \ref{eqn:dust}). Consequently, the timescale is strongly degenerate with respect to the initial surface density in planetesimals ($\Sigma_\s{pl, 0}$), the level of turbulence ($\beta_\s{1}$) and the radial concentration of the dust ($\beta_\s{2}$). 

Even a less massive planetesimal disk could produce indestructible seeds if the level of turbulence is correspondingly reduced or the radial concentration of the dust is increased. However, the final mass of the core will still ultimately be determined by $\Sigma_\s{pl, 0}$ (which sets the \textit{total} mass in solids available to be accreted). 

\section{Summary and discussion}\label{sec:conc}

In-situ formation models  \citep[e.g.][]{Hansen12, Chiang12} have recently become \textit{en vogue}, bolstered by the discovery of the large, close-in Kepler exoplanets and the ostensible inability of population synthesis models to reproduce the planets' observed properties. In the same spirit, we presented here a model that side-steps the km-size bottleneck in circumbinary configurations, allowing planetesimal accretion to proceed near a pressure trap. We identified a potential trap imposed by the density gradient inversion near the central binary, and postulated that solid bodies will tend to drift in and stop at that radius. We then sketched a possible scenario that takes into consideration planetesimal grinding, drift and reaccumulation.

We validated this scenario using a combination of 2D hydrodynamical and N-body simulations. We first showed that radial drift can be very fast for both small debris and km-sized planetesimals, and the pressure structure in the disk tends to accumulate both close to the central binary. Subsequently, we modeled the destruction-drift-reaccumulation process using a simplified numerical scheme, informed by our numerical simulations where possible. Within this model, we showed that if a population of primordial km-sized planetesimals is formed throughout the disk, a large fraction will be destroyed, the resulting debris accumulating close to the pressure trap. Surviving planetesimals will also drift into the trap, finding a solid-rich environment. A few seeds will grow large enough to become indestructible, start to accrete other planetesimals in the belt and entering a runaway growth phase.

Our toy model does not describe the evolution of the system beyond this critical size, since the large seeds will start stirring the planetesimal belt and the velocity dispersion will no longer be set by the binary/disk system alone. A full $N$-body simulation will be warranted to proceed further. However, the largest seeds should end up consolidating into a single, small core, which can then accrete the rest of the solid inventory in the belt (dust and planetesimals). In this scenario, the entire solid content within 3 AU is available for accretion by the planetary core. 

Is there enough solid material to form a core  matching the core masses inferred for the circumbinary Kepler planets? Again, Kepler-16 b represents the most stringest test among the observed circumbinary planets.The core of Kepler-16 b is expect to contain 40 to 60 Earth masses in heavy elements \citep{Doyle11}, despite the low metallicity of the central binary ([Fe/H] = $-0.3 \pm 0.2$). Assuming the standard MMSN normalization (10 g cm$^{-2}$), there are only about 3 Earth masses in heavy elements between 1 and 3 AU.  However, for close-in Kepler planets this normalization is likely underestimating the solid inventory. In particular, \citet{Chiang12} derived a ``minimum-mass extrasolar nebula'' (MMEN) assuming the close-in Kepler planets were formed in situ. The resulting normalization at 1 AU is approximately 50 g cm$^{-2}$. Accounting for some inefficiency in the planet formation process, then there might be just enough mass in solids to form the core of Kepler-16 b. This need for an enhanced normalization of the solid disk is common to all in-situ formation models \citep{Hansen12, Chiang12}.

We also remark that our model, for a given set of binary orbital elements, predicts a range of radii for the pressure trap, where we expect the planet to form. For the Kepler-16 binary parameters and nominal values for the disk parameters, the distance of the trap from the central binary is close to the current observed locations of the planet. To explain any discrepancy between the observed location of the planet and the position of the trap ($\approx .3$ AU for Kepler-16 in this paper), we can simply deviate from the nominal scale height normalization $h$ considered in this paper and fine-tune it to match the two radii (Equation \ref{eqn:drag}). Another possibility is to change the $\alpha$ parameter that determines the viscosity of the disk. The option of fine-tuning the radial location of the trap by varying the physical parameters of the disk slightly diminishes the determinism inherent to our model, so that we cannot uniquely predict (or postdict) the location of the planet given the binary parameters. Unfortunately, models in which the core formed far out and subsequently migrated in can also make analogous predictions, provided that $h$ and $\alpha$ are appropriately adjusted \citep[e.g.,][]{Pierens13}.  This is due to the fact that migrating cores will also stall close to the truncation edge \citep{Pierens07}.

The current scenario for circumbinary planet formation (formation of the core far from the binary, with subsequent migration) remains the most likely explanation for the origin of Kepler-16 b and the other circumbinary planets. Its main requirement is the existence of a migration mechanism (Type-I migration) which we \textit{already} know is at work within protoplanetary disks \citep[e.g.,][]{Kley12} and has shaped the planetary census. On the other hand, the alternative scenario presented in this paper makes a number of requirements (primarily, the existence of a primordial population of planetesimals to grind and the capture of both debris and lucky planetesimals close to the binary) in order to achieve the outcome of planet formation. An improved simulation addressing the crude simplifications and caveats listed in the next Section will help shed light on the viability of the model.

\subsection{Caveats and future work}\label{sec:caveats}
As mentioned throughout the paper, our approach required a number of simplifying assumptions in order to attempt to capture the physical processes at work within the computational constraints. Not all of these assumptions are fully self-consistent, but were chosen for the sake of simplicity and to allow the modeling of the system over $10^5$ years. In this respect, given the crude approximations mentioned in Section \ref{sec:results}, the scheme for planetesimal evolution presented here should be considered an ``order of magnitude'' approach to modeling the coupled evolution of the disk, destruction, drift and debris accretion on stalled planetesimals. This mainly allows us to derive a range of plausible timescales over which this process can be at work to produce a core. Only a more sophisticated treatment will verify whether some (or all) of the approximations detailed throughout the paper significantly alter, or even completely inhibit, the process described here. An $N$-body approach in the manner of P12 would lift some of the limitations of our model, while being able to model the collisional dynamics far more accurately. Such an improved simulation should try to address the following issues:

\textit{Planetesimal and debris sizes.} The limitation of a single-sized initial population is dictated by numerical convenience. Indeed, if we were to allow for a spectrum of sizes throughout the disk, then the numerical scheme would need to track planetesimal size evolution everywhere and lose much of its simplicity. Similarly, we do not follow the debris evolution in a detailed manner, but rather we lump them regardless of their size and categorize it as ``dust'' and assume it is instantaneously shepherded into the pressure maximum. We will lift this limitation in a future work (at the expense of increased computational burden). 

As mentioned in Section \ref{sec:dustaccr}, we calculate $\bar\epsilon$ (the dust accretion efficiency) by averaging over the possible encounter speeds $\Delta v_d$ between the planetesimals and the dust, assuming that $\epsilon = 1$ if $\Delta v_d < 100 v_{esc}$ and 0 otherwise. The bulk of the encounter speed arises from the mismatch between the gas eccentricity and the planetesimal eccentricity. For the small planetesimal size considered here (5 km), initially $\bar\epsilon(R = \mathrm{5 km}) \approx 0.1$, meaning that a majority of encounters would happen at high speeds ($\Delta v_d > 100 v_{esc}$). Our averaging implicitly assumes that such high-speed collisions do \emph{not} affect the target planetesimals (i.e. the dust flows around the planetesimal, or bounces without cratering). This is potentially the most problematic of the assumptions implicit in our model, considering that the high-speed collisions might ``sandblast'' the planetesimal instead. We note that only a detailed treatment of the accretion of dust of different sizes on the planetesimal (considering the balance between cratering and accretion) can give a more definite value for $\bar\epsilon$. Only for bigger planetesimals ($\radius \gtrsim 30$ km) the escape velocity is large enough that $\epsilon$ is always equal to 1 ($\Delta v_d$ is always less than $100 v_{esc}).$.

We also note that we measured $v_\s{coll}$ (the planetesimal collision speed throughout the disk, which determines the balance between the amount of dust produced and the number of planetesimals that survive long enough to drift into the belt) and $v_\s{drift}$ from an evolved distribution of planetesimals, so that any initial transients due to eccentricity oscillations have evolved towards an equilibrium profile. This essentially implies that the planetesimals would be born with the equilibrium orbital elements ``ab initio''.

\textit{Disk eccentricity and high-$m$, non-axisymmetric perturbations.} For simplicity, we chose to neglect the non-axisymmetric perturbations from the disk beyond the simple bulk disk eccentricity. These time-dependent perturbations could potentially disturb the formation of the planetesimal and dust rings.  However, we remark that in our hydrodynamical simulation the amplitude of the $m > 1$ modes is at least an order of magnitude smaller than the eccentric mode at 1 AU. 

As noted in Section \ref{sec:model_setup}, the assumption that each hydrodynamical quantity is constant along eccentric streamlines is strictly inconsistent with a non-constant disk eccentricity \citep{Statler01}. Indeed, we find that throughout the hydrodynamical runs, there is substantial scatter of $\Sigma$ at each semi-major axis. This issue is especially problematic close to 1 AU, where it is crucial to correctly model the dynamics of planetesimals in order to ascertain that planetesimals are captured close to the pressure maximum. We note that the results of our simulations are bolstered by the self-consistent simulation of \citet{Marzari08}, which found that small bodies can drift and congregate into a ring despite the disk perturbations.

Ascertaining the full impact of both approximations will require expensive hydrodynamical simulation fully coupled with the $N$-body code,  which are beyond the scope of the present paper.
 
\textit{Disk thermodynamics.} We choose to model the disk with an isothermal equation of state. While this choice greatly simplifies our equations, \citet{Marzari13} showed that the disk thermodynamics can be crucial in determining planetesimal collision speeds and orbital elements. Addressing this important shortcoming will require a substantial modification of our model.
 
 \textit{Multiple-planet systems?} We did not address the issue of multiple planet formation in this work. Thus far, we have only detected one such system, Kepler-47, consisting of two planets with $\mass_b \approx 7-10$ and $\mass_c \sim 16-23$ Earth masses, respectively. The inner planet is also relatively further away from the instability region than other circumbinary planets. The presence of the second planet poses the question of whether it could also form from a pressure trap, perhaps due to a density gradient induced by the formation of the innermost planet.  However, it is difficult to model this setup within the constraints of our model. This system will deserve further study to ascertain whether its observed properties can be explained within our framework, once some of its key assumptions are relaxed.

\textit{Impact of the fictitious eccentricity?} In our model and previous papers in the literature, both planetesimals and fluid elements are initialized in a circular orbit assuming a $G\mass_*/R$ potential, consistently with previous simulations. This was shown by \citet{Rafikov13} to introduce a small fictitious eccentricity in the initial conditions (such that both the disk and the planetesimals are initialized in a slightly eccentric state, especially close to the binary). In our case, this means that both the disk and the planetesimals are initialized with some initial eccentricity, potentially increasing both drift speeds and collision speeds. While this effect should be small compared to the eccentricities excited by the central binary, we will ascertain the impact of changing the initial conditions in a future $N$-body simulations.

\textit{Disk potential and self-gravity.} \citet{Rafikov13} also showed that including the contribution of the disk potential can have profound effects on both the planetesimal and binary dynamics. In particular, the eccentricity of planetesimals might be suppressed, moving the edge of the accretion-friendly region for Kepler-16 inwards down to about 2 AU. This effect could potentially affect our results as well, since \citet{Rafikov13} argues that fast binary precession could suppress the development of eccentricity in the disk. 

Further, including the self-gravity of the disk can in turn also affect both the disk and the planetesimal dynamics \citep{Marzari08}. Both effects are likely important ingredients needed to capture the full dynamical picture, but are neglected for simplicity in our model. It is not necessarily clear 

\textit{Snow line, layered structure and turbulence.} Our simplified, two-dimensional model does not take into account any stratification or structure, assuming that the whole disk can be represented with a simple $\alpha$-disk model. In reality, the snow line can also act as a solid trap due to the pressure maximum there; therefore, the inner disk will be ``shielded'' by the snow line, and planetesimal grinding in the outer disk will not result in migration of debris in the inner disk. We have attempted to crudely capture this effect by only considering dust production within the inner 3 AU. 

\citet{Martin13} consider layered circumbinary disk models, and find that a dead zone could likely extend from the inner edge of the disk to several AUs and remain relatively unaltered for the disk lifetime. Within the dead zone, solids could drift towards a peak in the surface density (located at a few AUs from the stellar binary in their models), and settle within a fairly quiescent midplane. A similar process as that outlined in the present paper could then still be in action, but concentrating the material further out. More sophisticated self-consistent simulations that include layering will be required to assess this possibility.

As addressed in \citet{Meschiari12c}, even within a dead zone planetesimals can be stochastically kicked by residual turbulent torques, raising their eccentricity and dephasing their orbits. Additionally, a radially concentrated ring of planetesimal will be smeared radially (increasing $\Delta R_\s{b}$ and reducing the collision rate in the planetesimal belt). Low levels of turbulence are therefore beneficial to our scenario. We plan to include turbulent torques in the manner of \citet{Meschiari12c} in future work.

\acknowledgments
S.M. acknowledges insightful discussions with Augusto Carballido, Sarah Dodson-Robinson, Greg Laughlin and Mike Pavel,  as well as support from the W. J. McDonald Postdoctoral Fellowship. The author would like to thank the anonymous referee for their extremely useful and rigorous critique. 

\bibliographystyle{apj}

\end{document}